\documentclass[
    ,final            
  ]
  {aipproc}

\layoutstyle{6x9}

\def\NJ{N_{\J}}
\def\J{J/\psi}
\def\Nc{N_c}
\def\Ncbar{N_{\bar c}}
\def\ccbar{c \bar c}
\def\Nccbar{N_{\ccbar}}

\newcommand{\AmS}{{\protect\the\textfont2
  A\kern-.1667em\lower.5ex\hbox{M}\kern-.125emS}}

\begin{document}

\title{Nonlinear Behavior of Quarkonium Formation and
Deconfinement Signals}

\author{R. L. Thews}{
  address={Department of Physics, University of Arizona, Tucson,
AZ 85718 USA}
}

\begin{abstract}
We anticipate new features of quarkonium production in heavy ion collisions
at RHIC and LHC energies which differ from a straightforward 
extrapolation of results at CERN SPS energy. General arguments indicate 
that one may expect quarkonium formation rates
to increase more rapidly with energy and centrality 
than the production rate of the heavy quarks which they
contain.  This is due to new formation mechanisms in which 
independently-produced quarks and antiquarks form a bound
quarkonium state.  This mechanism will depend quadratically on the
total number of initially-produced heavy quark pairs, and becomes
numerically significant only at RHIC and LHC energy.  When viewed as
a signal of color deconfinement, a transition from suppression to enhancement
may be observed.  Explicit model calculations are presented, in which
one can follow striking variations of final quarkonium production
within a range of parameter space.
\end{abstract}

\maketitle


\section{Introduction}

The production of heavy quarkonium states in high energy hadronic interactions
proceeds through creation of the corresponding flavor heavy quark-antiquark
pair.  Given the heavy quark mass to provide a perturbative scale, one
can employ a perturbative QCD calculation for this initial process
\cite{altarelli}.
The spectrum of 
heavy quarkonium states can be described by essentially non-relativistic 
heavy quarks interacting via a static potential.  Phenomenological 
extraction of the potential leads to a linear rise at large separation,
which provides the observed quark confinement. This potential can also
be directly calculated via lattice methods \cite{karsch}, 
and confirms these general
properties.  These lattice methods can also be utilized for QCD at finite
temperature, which reveals that at sufficiently high T the QCD spectrum 
will change from confined hadrons to colored degrees of freedom.  
The corresponding heavy quark static potential in this case shows
a decrease in the long range part, and disappears above the deconfinement
temperature.  The goal of high energy heavy ion collision experiments is
to create a region of space-time within which these finite temperature
predictions can be tested.

A signature of color deconfinement which utilizes heavy quarkonium
production rates was proposed more than 15 years ago \cite{matsuisatz}.  
One invokes the argument that in the deconfined region where the 
color-confining force has disappeared,  a
heavy quark and antiquark
cannot form a quarkonium bound state, 
and they may diffuse away from each other to separations larger than
typical hadronic dimensions.
As the system cools and the confining potential reappears, these heavy
quarks will not be able to ``find'' each other and form heavy
quarkonium.  They will then bind with quarks which are close by to them
at hadronization.  Since these quarks are predominantly the lighter
u, d, and s flavors, they will most likely form a final hadronic state with
``open'' heavy flavor.
The result will be a
decreased population of heavy quarkonium relative to those which would
have formed if a region of deconfinement had not been present.  This
scenario as applied to the charm sector is known as $\J$ suppression.
There are now extensive data on $\J$ production
using nuclear
targets and beams.
Results from the NA50 experiment at the CERN SPS reveal an  
``anomalous"
suppression, prompting claims that
this effect could be the expected signature of deconfinement \cite{NA50}.
Measurements at higher energies of RHIC, and eventually
at the heavy ion runs at LHC, should be able to provide enough information
to either support or refute these claims.  
Straightforward extension of the deconfinement scenarios to these
higher energies anticipate that
$J/\psi$ suppression would be virtually complete at all
centralities \cite{Vogt}.

The purpose of this work is to point out that there will be
another consequence of the increased beam energy for the suppression
scenario.  This is because one expects that multiple pairs of 
charm-anticharm quarks will be produced in the initial partonic
stage of the collision.  Perturbative QCD estimates predict about 
10 charm pairs at RHIC energy, and several hundred pairs 
at LHC \cite{hardprobes1}.  
This situation provides a ``loophole'' in the Matsui-Satz 
argument, since there will be many more heavy quarks in the 
interaction region with which to combine.  In order for this to
happen, however, one must invoke a physical situation in which
quarkonium states can be formed from {\em all combinations} 
of the heavy quark pairs.  If this is possible, then the rate of 
quarkonium formation from $N$ initially-produced quark-antiquark
pairs would initially be proportional to $N^2$.  If the 
total number of quarkonium states remains a small fraction of
the total number of pairs, then the final population will retain
this quadratic dependence.  Although still small compared with $N$, this
number can be much larger than the ``ordinary'' expectations which
are linear in $N$.

In the next section, we go through the pQCD methods and results
which lead to the large $N$-values at high energy.  The variation of
these quantities with centrality is also explored.  The following section
presents generic arguments for the properties of
quarkonium formation in any physical situation for which the 
quadratic process is allowed.  The last two sections present 
results specific to two physical models which share 
some of these properties.

\section{Heavy Quark Production in A-A Collisions}

The calculation of heavy quark production by hadrons is based on 
perturbative QCD processes at the parton level.  The perturbative
approach requires a large scale to justify the expansion in powers 
of $\alpha_s$, which is provided by the heavy quark mass.  One then uses 
hadronic structure functions measured in other reactions (e.g. Deep 
Inelastic Scattering and Drell-Yan) plus factorization to calculate
the hadronic cross section.  The general expression is of the form

\begin{equation}
\sigma(s,m_Q,\mu_F,\mu_R) = \sum_{i,j}\int dx_1 \int dx_2 F_i(x_1,\mu_F)
F_j(x_2,\mu_F) {\hat{\sigma}}_{ij}(\hat{s},m_Q,\mu_R)
\end{equation}

where $i$ and $j$ label the initial state partons, F(x,$\mu_F$) are the
structure functions, evaluated at a factorization scale $\mu_F$, and
${\hat{\sigma}}_{ij}$ are the partonic cross sections for
producing a heavy quark-antiquark pair, which depend on the partonic process
subenergy $\hat{s} = x_1 x_2 s$, the heavy quark mass $m_Q$ and the 
renormalization scale $\mu_R$.

The partonic cross sections have been calculated 
\cite{nde1},\cite{mnr1},\cite{mnr2} to lowest order 
(LO), in which quark-antiquark annihilation and gluon-gluon fusion and
annihilation into $Q\bar Q$ contribute to order ${\alpha_s}^2(\mu_R)$, and
also in next to leading order (NLO), where real gluons emitted in the
final states of the LO processes, plus processes involving quark or
antiquark plus gluon in the initial state, plus virtual loop corrections
all contribute to order ${\alpha_s}^3(\mu_R)$.  There is additional 
dependence on $\mu_R$ in NLO, in the form of terms proportional to
$ln({\mu_R^2}/{m_Q^2}).$  In practice one usually takes $\mu_R = \mu_F \equiv
 \mu$, 
but in general they are independent parameters of the calculation,
along with $m_Q$. In addition, there are several sets of parton distribution
functions which can be utilized.  In order to constrain these 
parameters, a comprehensive comparison with existing charm production
data in N-N and $\pi$-N reactions was undertaken \cite{hardprobes1}.
For a recent update of this procedure, see Reference \cite{vogtnew}.

\begin{figure}[t]
\includegraphics[clip=,height=.55\textheight]{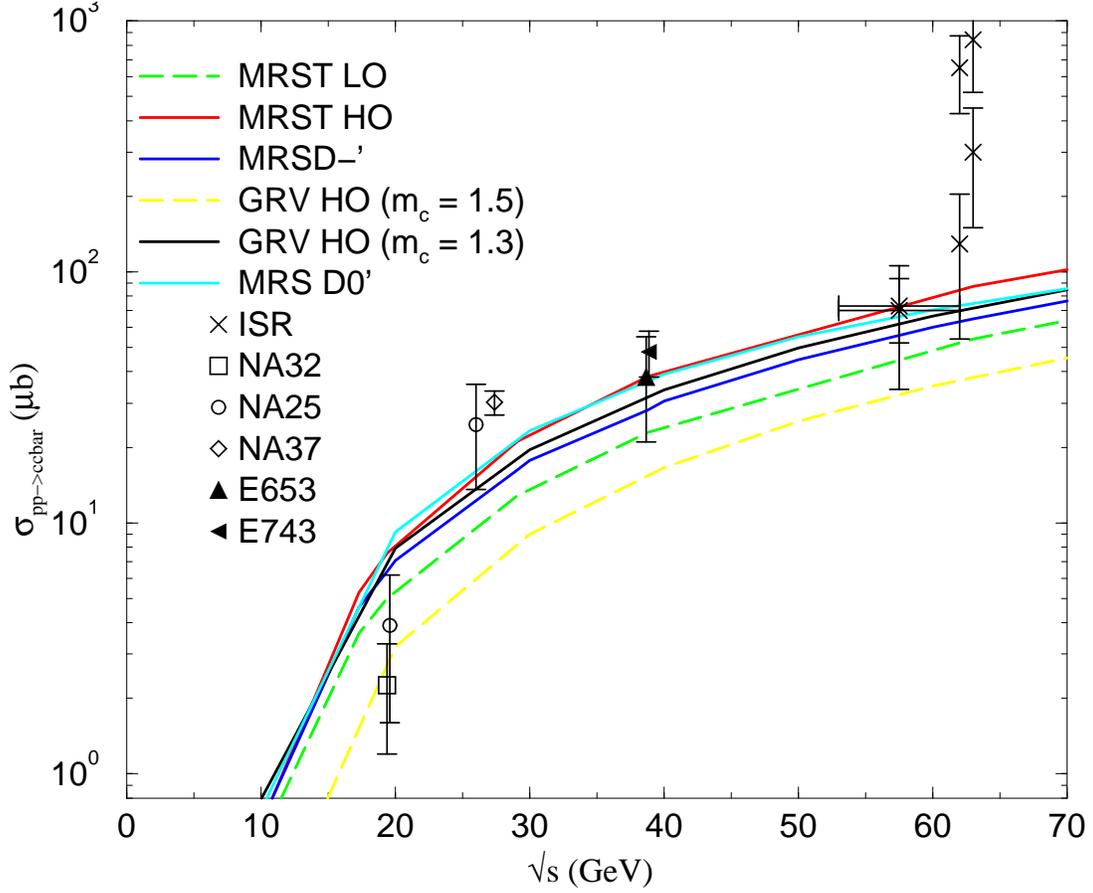}
\caption{\small pQCD calculated cross section for $pp \rightarrow c\bar c$.}
\label{sigmacharmlow}
\end{figure}

The charm mass $m_c$ was allowed to vary between 1.2 and 1.5 GeV, and
$\mu$ was varied between $m_c$ and 2$m_c$.  The results for the
cross section $pp \rightarrow c\bar c$ in the energy interval
10 GeV < $\sqrt{s}$ < 70 GeV are shown in Figure \ref{sigmacharmlow}.

Results from the original structure function sets MRSD0 and MRSD-$^{\prime}$ 
are shown, and supplemented by the more recent sets of structure
functions MRST\,HO and GRV\,HO.  The difference in predictions between these
sets is in general smaller than the experimental uncertainties, except
for the GRV with large $m_c$ and the low order set MRST\,LO, which is
only shown for comparison (for consistency one must use the NLO version
of the structure functions together with NLO partonic cross sections).
However, we must remember that the x-values probed in these calculations
are always greater than $m_c/\sqrt{s}$, which is large enough such that
all structure functions are very well constrained by the DIS and DY data.
What one may worry about is the large change in cross section between the LO
alone and the total LO + NLO.  The ratio of these values (the K-factor)
typically varies between 1.5 and 2.5 in this energy region.  Thus one 
might expect that higher order terms in the perturbative expansion (NNLO)
might also be as significant as the NLO.  In this case, a satisfactory 
fit to the data would probably require different values of the mass
and scale parameters, thus altering the predictions at higher energy.

\begin{figure}[t]
\includegraphics[clip=,height=.56\textheight]{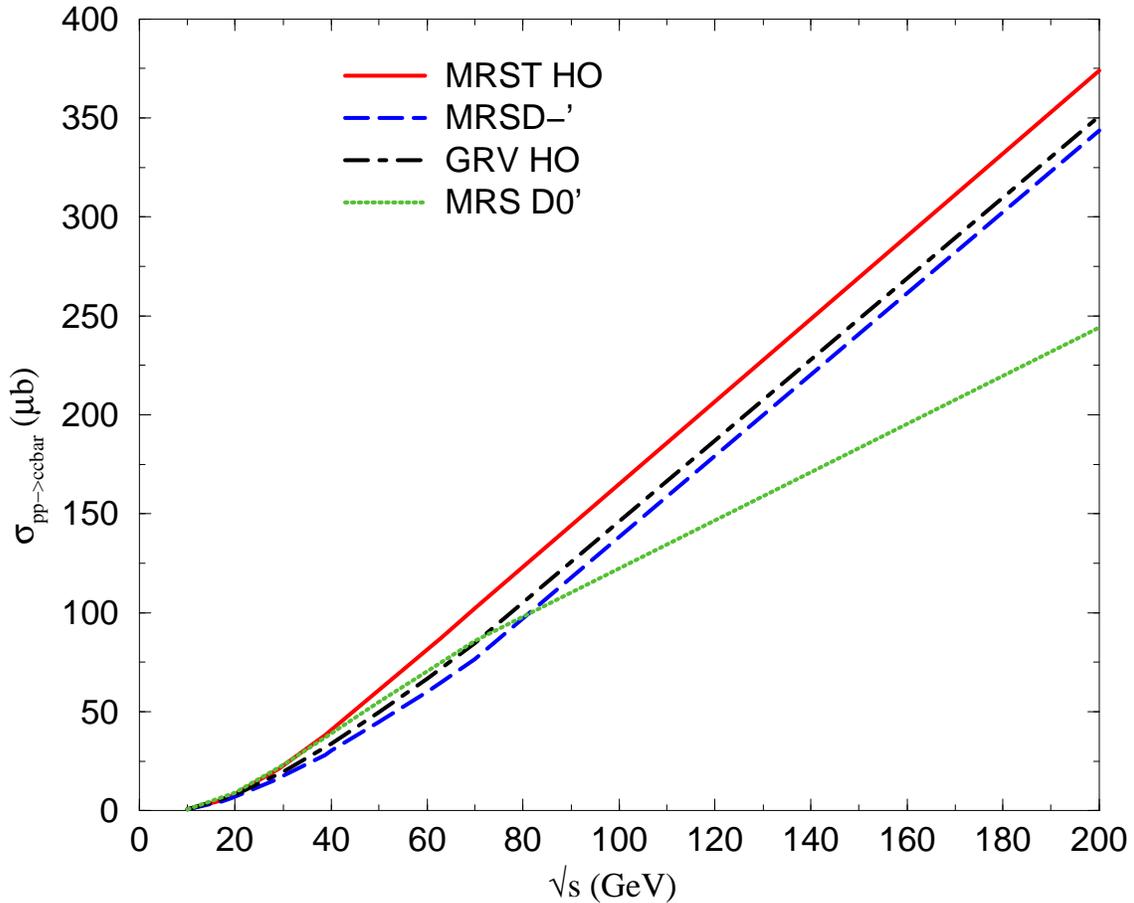}
\caption{\small pQCD calculated cross section for $pp \rightarrow c\bar c$
extrapolated to RHIC energy.}
\label{sigmacharmrhic}
\end{figure}

The extrapolation of these calculations up to RHIC energies is shown
in Figure \ref{sigmacharmrhic}.  
We show only the four structure function
sets which agree with the low energy data.  One sees that there is
some divergence at the highest $\sqrt{s}$ = 200 GeV.  However, this plot
is on a linear scale, which maximizes the appearance of the differences.

\begin{figure}[t]
\includegraphics[clip=,height=.52\textheight]{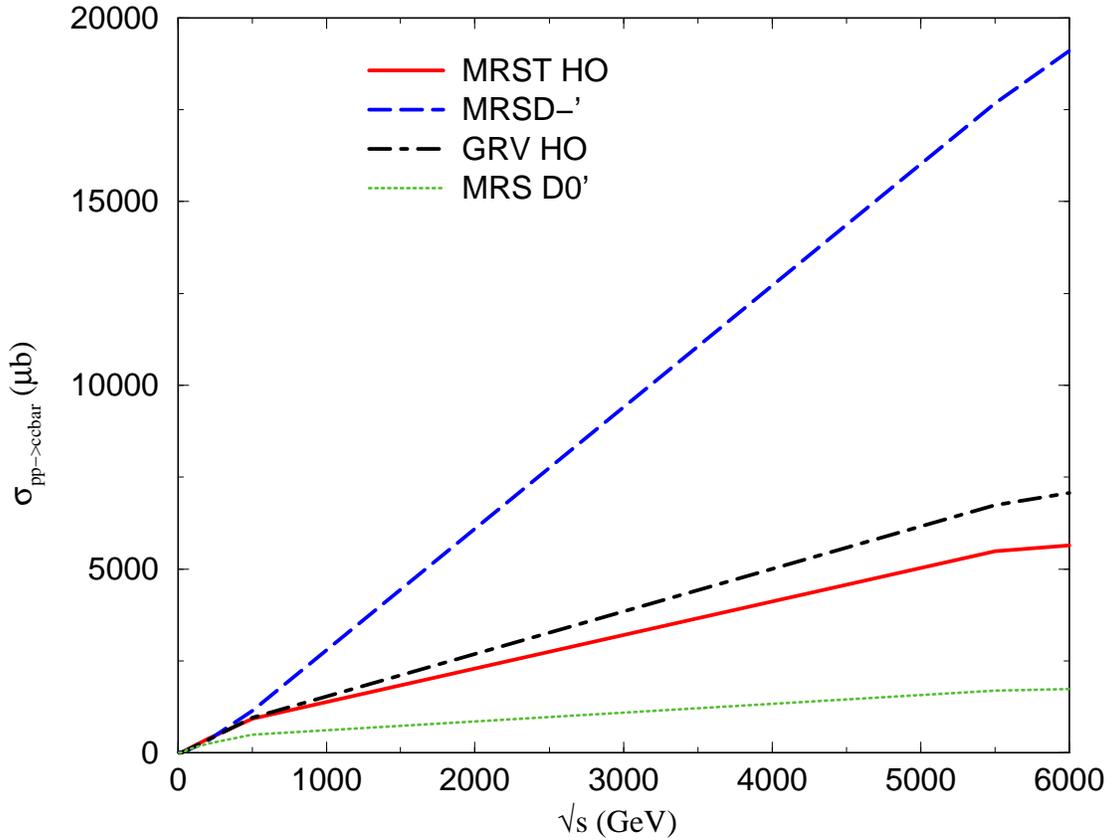}
\caption{\small pQCD calculated cross section for $pp \rightarrow c\bar c$
extrapolated to LHC energy.}
\label{sigmacharmlhc}
\end{figure}

Finally, we show in Figure \ref{sigmacharmlhc} the extrapolation of
these calculations to the LHC heavy ion energy region. 
The scale is again linear, and there is almost a factor of 10 difference
between the old structure function sets.  This can be attributed to the
low x-values probed at $\sqrt{s}$ = 5500 GeV, down to about
$x \approx 3 \times 10^{-3}$.  The more recent structure functions
include DIS data from HERA which 
is sensitive to these low x-values, and we
see that both the MRST and GRV HO sets follow each other much more closely.
Note however, that we have not used the complete set of modern structure 
functions and parameter space values which fit the low energy data.
The calculations in Reference \cite{vogtnew} which do include all
of these possibilities predict a range which differs by a factor of 2.3
between highest and lowest values.

We now use these cross sections to predict the number of heavy quark
pairs which will be produced in heavy ion collisions.  In the simplest
case when we assume that the heavy ion collision is just an incoherent
superposition of N-N collisions (no shadowing corrections), we only
need calculate the integrated N-N luminosity in a single A-A collision.
For a central collision between identical nuclei, geometry tells us
this should be of the form $A^2/\pi R_A^2 \propto A^{4\over 3}$.

For a more general calculation, one needs a parameterization
of the nuclear density $\rho_A(\vec{s},z)$, where z is the coordinate
along the beam direction and $\vec{s}$ is the 2-dimensional
position vector in the transverse plane.  Then one can calculate
a nuclear thickness function

\begin{equation}
T_A(\vec{s}) = \int dz\, \rho_A(\vec{s},z)
\end{equation}

which is normalized to the total nuclear number

\begin{equation}
\int d^2\,\vec{s}\: T_A(\vec{s}) = A.
\end{equation}

Then consider a collision of two nuclei which are incident along 
paths parallel to the z-axis separated in the
transverse plane by a vector $\vec{b}$ (the magnitude of $\vec{b}$
is the impact parameter b).  The integrated N-N luminosity is then
just the product of the two nuclear thickness  functions integrated
over each overlap point in the transverse plane.  This is called
the nuclear overlap function:

\begin{equation}
\int d^2\,\vec{s}\; T_A(\vec{s})\, T_B(\vec{b}-\vec{s}) \equiv T_{AB}(b)
\end{equation}
which by axial symmetry can only depend on the impact parameter b.

\begin{figure}[p]
\includegraphics[clip=,height=.44\textheight]{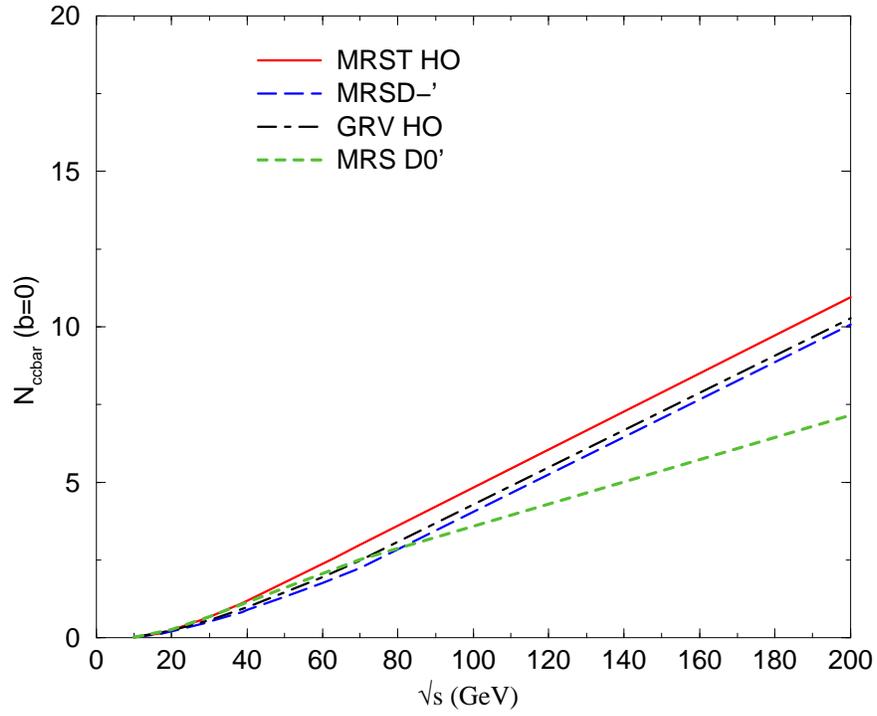}
\caption{\small Energy dependence of central $\Nccbar$ in RHIC region.}
\label{ncharmrhic}
\end{figure}

\begin{figure}[p]
\includegraphics[clip=,height=.44\textheight]{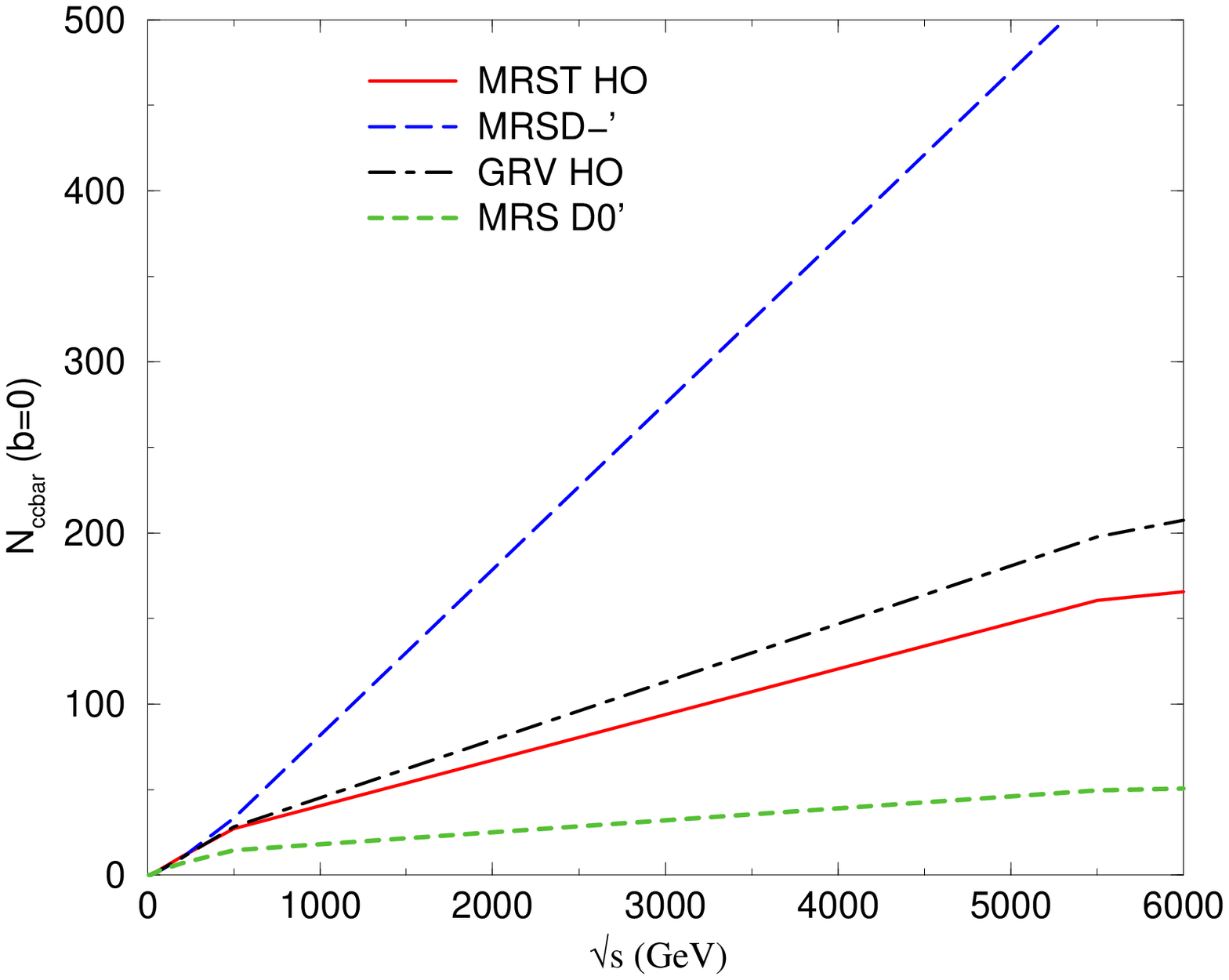}
\caption{\small Energy dependence of central $\Nccbar$ in LHC region.}
\label{ncharmlhcion}
\end{figure}

Calculations using standard nuclear density profiles produce a typical value 
for heavy ion (e.g. Pb-Pb or Au-Au) of $T_{AA}(b=0) \approx 30 mb^{-1}.$
This leads to the estimates of $\Nccbar$ for central collisions at the
various experimental facilities.  Taking average values for the 
open charm cross section calculations, we obtain $\Nccbar$(b=0) = 
0.2 (SPS), 10 (RHIC200), and 200 (LHC).  The energy dependence and
variation with structure function set are shown in Figures \ref{ncharmrhic} 
and \ref{ncharmlhcion}.

The first RHIC measurement of open charm has been reported at this
institute \cite{zajc},\cite{phenix}.  From observation of 
high-$p_T$ electrons by the PHENIX Collaboration from the 130 GeV run, 
an equivalent value of the p-p open charm cross section can be
extracted.  For central collisions, the reported value is
$\sigma(pp\rightarrow c\bar c) = 380 \pm 200 \mu b$.  
Although the uncertainties are still quite large, we can already
draw some conclusions and make some interesting speculations.

\begin{figure}[tb]
\includegraphics[clip=,height=.50\textheight]{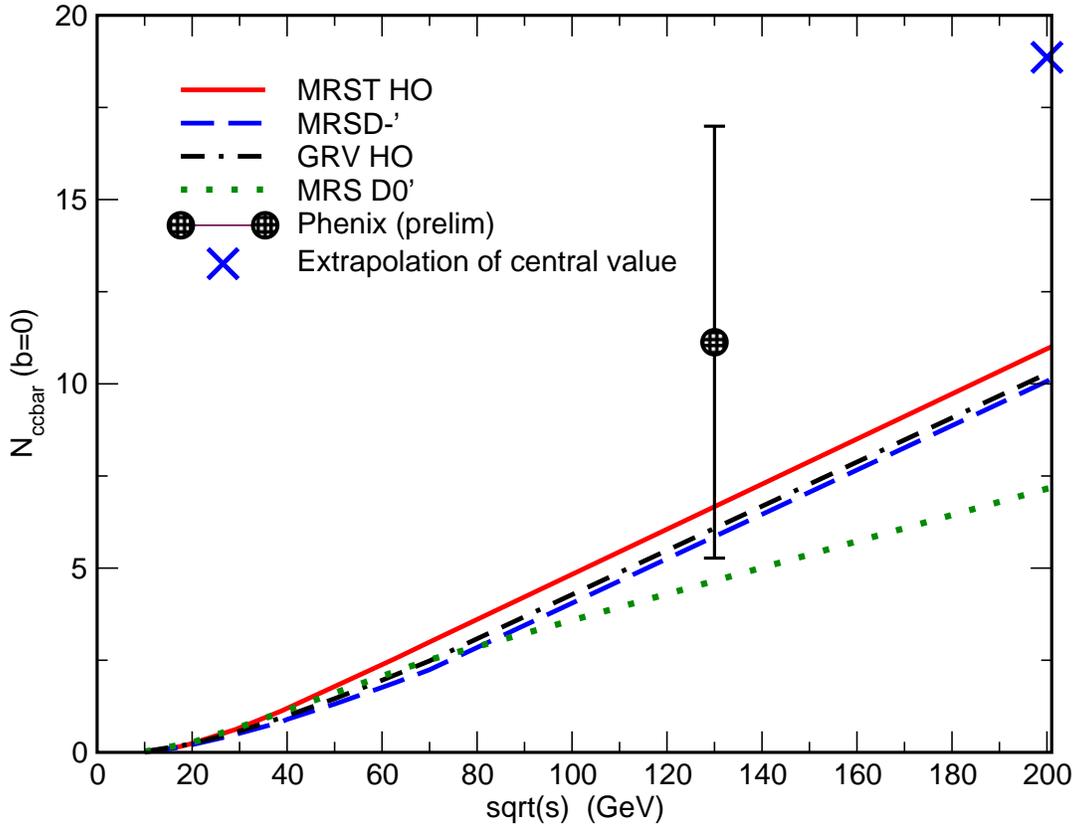}
\caption{\small Comparison of PHENIX measurement with pQCD calculations of
central $\Nccbar$.}
\label{ncharmphenix}
\end{figure}

The corresponding $\Nccbar$ values for b = 0 collisions are overlayed 
on Figure \ref{ncharmrhic} and shown in Figure \ref{ncharmphenix}.
One sees that the magnitude of the calculated values are consistent
with the measurement within errors.  However, the central value is 
well above the calculated values, even exceeding the nominal prediction 
at the higher 200 GeV energy.  A simple extrapolation of this 
central point to $\sqrt{s}$ = 200 GeV would imply $\Nccbar$ between
15 and 20.  This is substantially above the nominal estimate of 10, and
could enhance the expected nonlinear effects for $\J$ formation.
This situation is somewhat surprising, since these values are extracted
just from nuclear geometry and calculations using structure functions
of nucleons.  One might expect that there will be a depletion
of gluons in a heavy nucleus relative to free nucleons, similar to
the shadowing of quark structure functions observed for DIS with
nuclear targets \cite{eks}.  These speculations, of course, are only
relevant assuming the eventual uncertainties of the measured value
will converge toward the current central value.

\begin{figure}[p]
\includegraphics[clip=,height=.40\textheight]{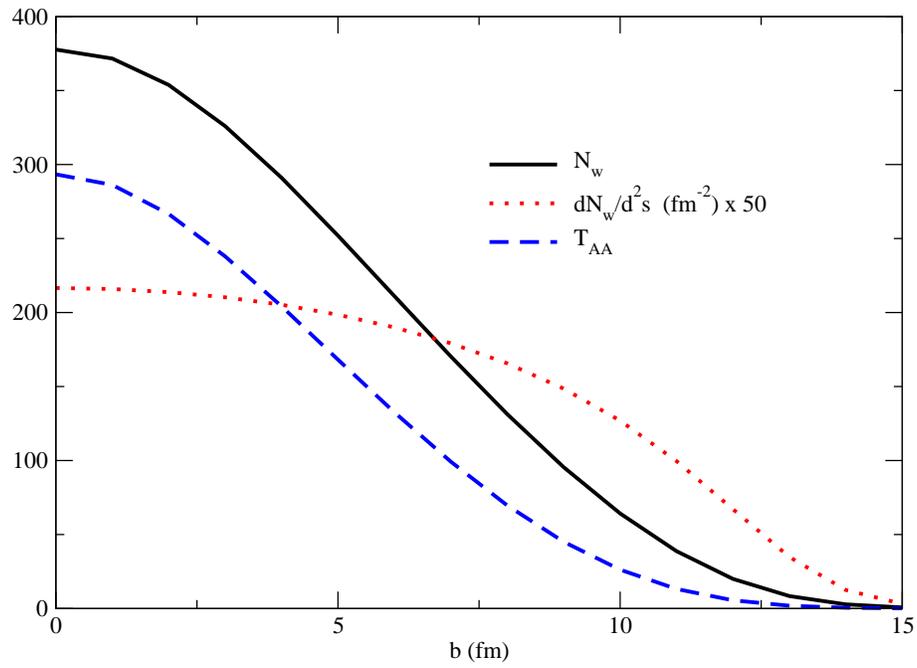}
\caption{\small Wounded nucleon and nuclear overlap dependence on impact parameter.}
\label{impactparameter}
\end{figure}

\begin{figure}[p]
\includegraphics[clip=,height=.49\textheight]{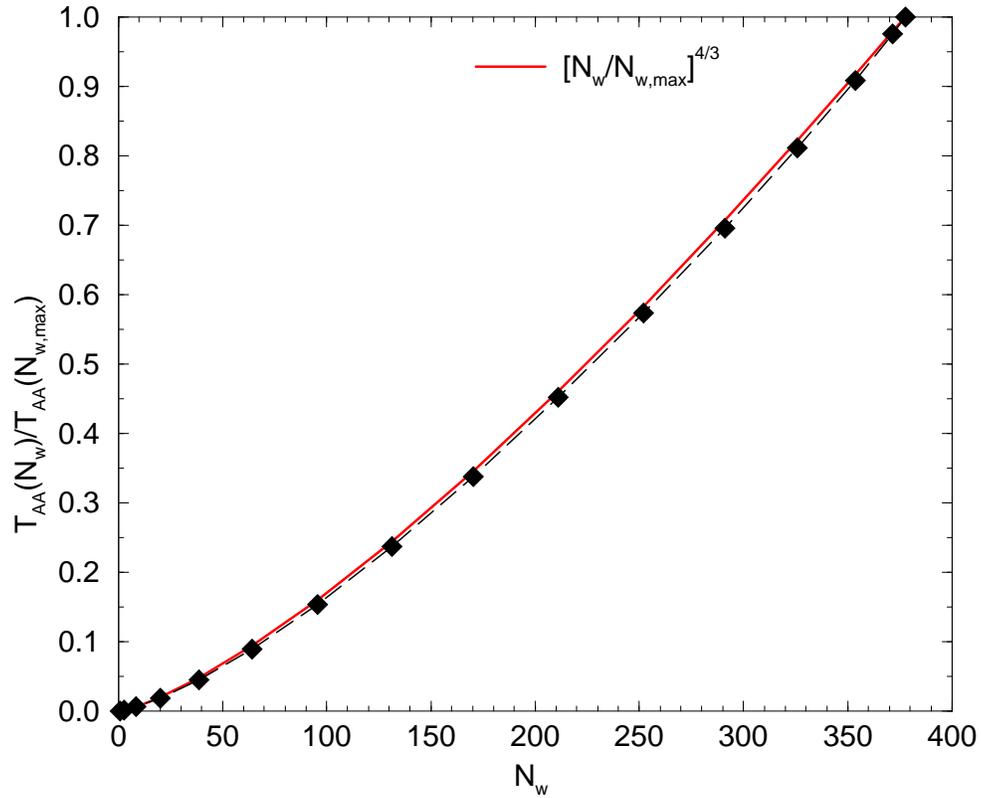}
\caption{\small Centrality dependence of $T_{AA}$ in terms of $N_w$.}
\label{taavsnw}
\end{figure}

We can also use the overlap function $T_{AA}(b)$ to predict the
centrality dependence.  However, the impact parameter $b$ is not directly
measurable.  Instead, the number of participant nucleons $N_p$ is
generally used as a connection between nuclear geometry and 
experimental measurables.  Most experiments measure the transverse
energy $E_T$ of each event and relate this to centrality.  
For this, one needs a model for $N_p$.  One popular choice is the
wounded nucleon model \cite{wounded}, in which every nucleon which
undergoes at least one inelastic collision is called ``wounded'' and
is counted as a participant nucleon.  The utility is reinforced by
the experimental observation that $N_p$ and $E_T$ are linearly-related over 
quite a wide range in centrality \cite{na49}.  We show in Figure 
\ref{impactparameter} the calculated dependence of $N_w$ on impact parameter, using standard nuclear density profiles and an inelastic N-N 
cross section of 50 mb.  A similar behavior to that of the nuclear overlap 
function $T_{AA}$ is evident.  
Also shown for future reference (dotted line) is the
wounded nucleon density in impact parameter space (evaluated at the
center of the overlap area).   
We can then recast the 
centrality dependence of $T_{AA}$(b)
in terms of the participant (or in this case wounded)
nucleon number.   This is shown in Figure \ref{taavsnw}. 

One sees that there is a power law dependence, $T_{AA} \propto N_w^{4/3}$.
This relationship could have been anticipated from the general arguments
concerning equivalent N-N integrated luminosity, but it is pleasing to
verify in an explicit model.
  

\begin{figure}[tb]
\includegraphics[height=.53\textheight]{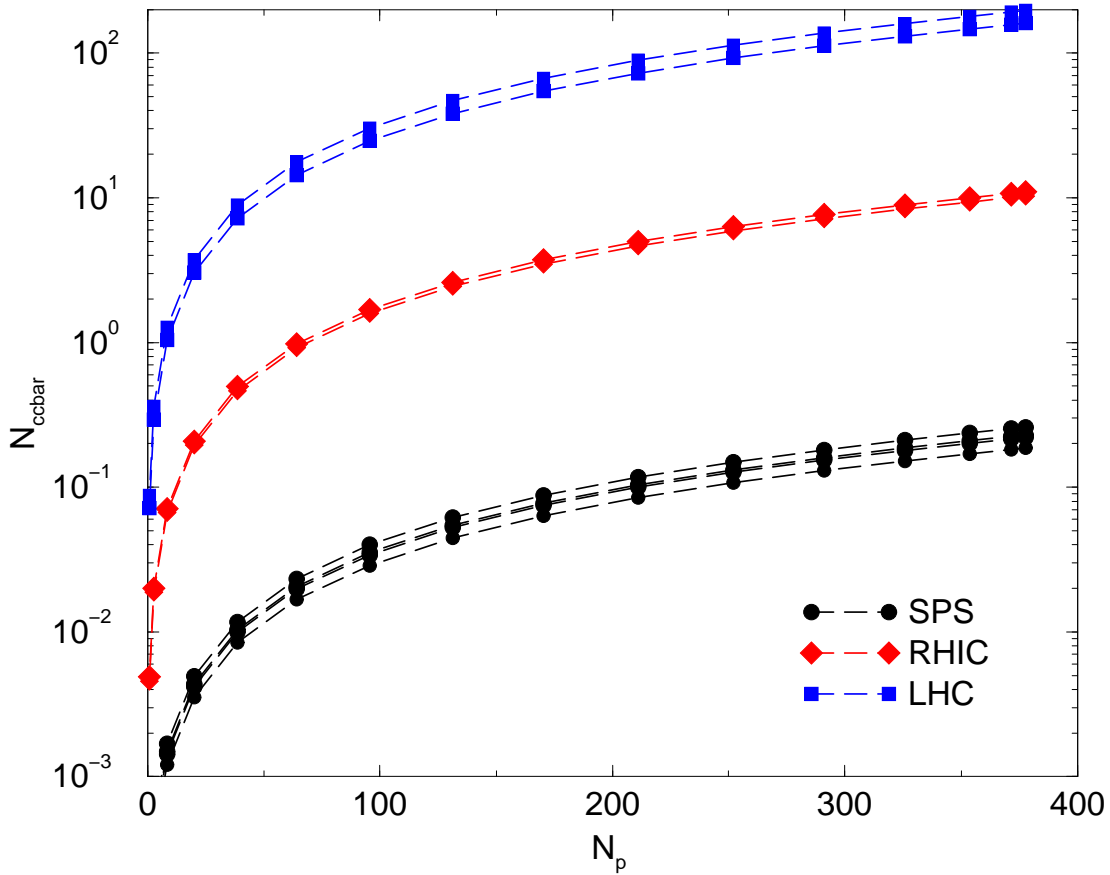}
\caption{\small Centrality dependence of $\Nccbar$.}
\label{ncharmvsnp}
\end{figure}

Finally, one can plot the predictions for $\Nccbar$ as a function of
centrality, using participant nucleon number $N_p$ as a label.
This is shown in Figure \ref{ncharmvsnp} for SPS, RHIC200, and
LHC energies.  The multiple points in each curve come from using
several structure function sets.  The decrease with centrality 
just follows the power law behavior of $T_{AA}$, and one sees
that at sufficiently peripheral collisions the average number
of charm pairs produced will decrease below unity for all energies
considered.  This behavior will be a very useful constraint on the 
models to be considered.

\section{Quarkonium Formation from Uncorrelated Pairs}

As shown in the previous section for
heavy ion collision energies at RHIC and LHC, the 
initial number of heavy quark pairs produced in each
collision will be qualitatively different than the number produced
at SPS energies.  To be specific, let us consider charm quarks and
the subsequent production/formation of $\J$.
Typically, the number of charm quark pairs is 
expected to be of the order of ten at RHIC and several hundred at LHC 
in the most central collisions \cite{hardprobes1}. Let us attempt to
extract features of $\J$ formation from the initially-produced charm
quarks which are independent of detailed dynamics \cite{thews3}.

We consider scenarios in which the formation of $\J$ is allowed to
proceed through any combination of one of the $\Nc$ charm quarks with
one of the $\Ncbar$ anticharm quarks which result from the initial production
of $\Nccbar$ pairs in a central heavy ion collision.  This of course 
would be expected to be valid in the case that a space-time region
of color deconfinement is present, but is not necessarily limited to
this possibility.  For a given charm quark, 
one expects then that the probability $\cal{P}$ to form a $\J$ is 
proportional
to the number of available anticharm quarks relative to the number of
light antiquarks, 
\begin{equation}
{\cal{P}} \propto \Ncbar / N_{\bar u, \bar d, \bar s} \approx \Nccbar / N_{ch}.
\end{equation}
In the second step we have replaced the number of available anticharm 
quarks by the total number of pairs initially produced, which assumes that
the total number of bound states formed remains a small fraction of the total.
Also, we normalize the number of light antiquarks by the number of 
produced charged hadrons.  
Since this probability  is generally very small, one can simply multiply by the
total number of charm quarks $\Nc$ to obtain the number of
$\J$ expected in a given event.  

\begin{equation}
\NJ \propto {\Nccbar}^2 / N_{ch},
\label{eqquadratic}
\end{equation}
where the use of the initial values $\Nccbar = \Nc = \Ncbar$ is again justified
by the relatively small number of bound states formed.
For an ensemble of events, the average number of $\J$ per event is calculated
from the average value of initial charm $<\Nccbar>$, and we neglect fluctuations
in $N_{ch}$.
\begin{equation}
<\J> = \lambda (<\Nccbar>+1)<\Nccbar> / N_{ch},
\label{eqgeneric}
\end{equation}
where we place all dynamical dependence in the parameter $\lambda$.

One can extend this formula to the case where $\J$ formation is effective
not over the entire rapidity range $Y_{total}$, but 
only if the quark and antiquark are within the same rapidity interval 
$\Delta y$. There is a significant simplification if the rapidity
dependence of the charm quark pairs and the charged hadrons (or 
equivalently the light antiquarks) are the same.
In this case the entire effect is just the replacement 
$<\Nccbar>+1 \rightarrow <\Nccbar> + Y_{total} / \Delta y$.  However,
one must remember that the prefactor $\lambda$ will in general
contain some dependence on the size of the rapidity interval.

The essential property of this result is that the growth with energy
of the term quadratic in total charm \cite{hardprobes1}
is expected to be much stronger than
the growth of total particle production in heavy ion collisions
\cite{ncharged}.  $\J$ production without this quadratic mechanism is
typically some small energy-independent fraction of total initial charm
production \cite{hardprobes2}, so that we can expect the quadratic
formation to become dominant at high energy.

We show numerical results in Table 1 for these quantities with a
prefactor $\lambda$ of unity.  Estimates for the charm and particle
numbers are very approximate, but serve to show the anticipated trend
with energy.  At SPS, this formation mechanism is most probably 
insignificant.  At RHIC it is comparable with ``normal" formation, while
at LHC one might expect it to be dominant.  Of course, the exact result
will depend on the details of the physics which controls the formation.

\begin{table}[tb]
\caption{Comparison of $\J$ formation variation with energy}
\label{table:1}
\newcommand{\m}{\hphantom{$-$}}
\newcommand{\cc}[1]{\multicolumn{1}{c}{#1}}
\renewcommand{\tabcolsep}{2pc} 
\renewcommand{\arraystretch}{1.2} 
\begin{tabular}{@{}llll}
\hline
           & \cc{$SPS$} & \cc{$RHIC$} & \cc{$LHC$}  \\
\hline
$\sqrt s$ (GeV)          & \m18 & \m200 & \m5500  \\
$<\Nccbar>$               & \m0.2 & \m10 & \m200 \\
$N_{ch}$                  & \m1350  & \m3250  & \m16500  \\
$<\NJ>$                   & \m0.00018 & \m0.034 & \m2.4 \\
$\NJ^{initial}$           & \m0.0012 & \m0.06 & \m1.2 \\
\hline
\end{tabular}\\[2pt]
\end{table}

We can also estimate the centrality dependence of $\J$ production from
Equation \ref{eqgeneric}.  The number of charm quark pairs should obey
\begin{equation}
\Nccbar \propto {N_p}^{4/3}
\end{equation}
from the properties of the nuclear overlap function.  The number of hadrons
produced generally scales with the number of participants,
\begin{equation}
N_{ch} \propto {N_p}^{\alpha},
\end{equation}
where one has measured values $\alpha = 1.07 \pm 0.04$ at SPS \cite{SPSref} and
$\alpha = 1.13 \pm 0.05$ at RHIC-130 \cite{RHICref}.  At LHC, one might
anticipate that hadron production would become dominated by QCD minijets
\cite{ncharged}, so that $\alpha \approx 4/3$. However, a comparison of RHIC
results at 130 and 200 GeV does not indicate a dramatic effect in this
energy range. \cite{PHOBOS200},\cite{BRAHMS200}

Given these values, one
can predict
\begin{equation}
\NJ \propto {N_p}^{\beta={8\over 3} - \alpha}
\end{equation}
for collisions in which $<\Nccbar> \;\; \gg 1$, and
\begin{equation}
\NJ \propto {N_p}^{\beta={4\over 3} - \alpha}
\end{equation}
for collisions in which $<\Nccbar> \;\; \ll 1$.

\begin{figure}[p]
\includegraphics[clip=,height=.44\textheight]{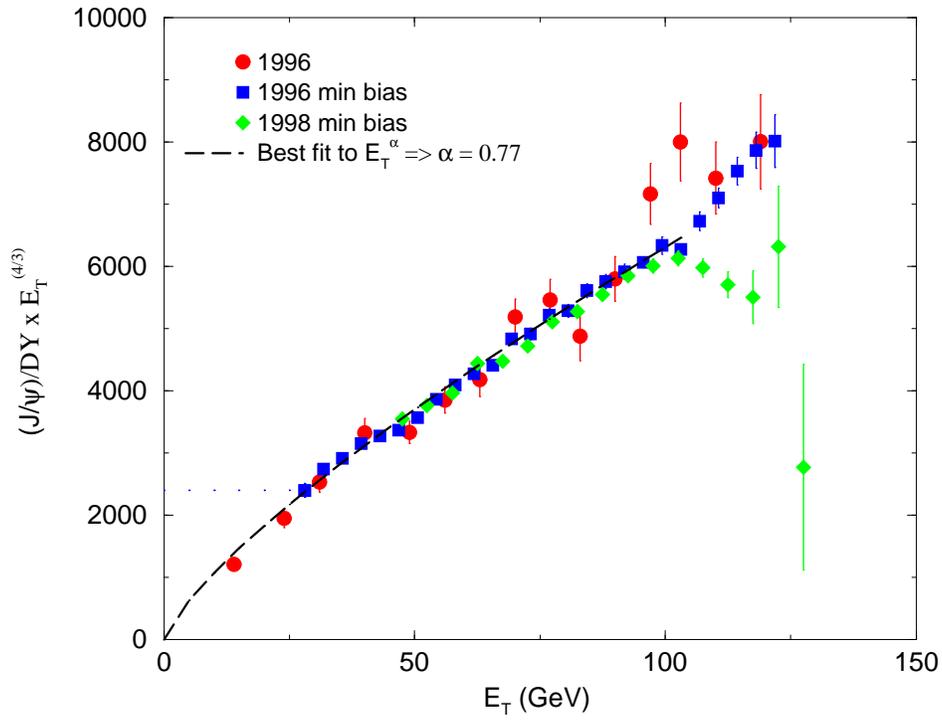}
\caption{\small Centrality dependence of $\J$ inferred from measured ratio
$\J$/Drell-Yan by NA50 at SPS.}
\label{NA50jpsioverDY}
\end{figure}

One can make an indirect check of the prediction at SPS energy, utilizing
the NA50 data \cite{NA50} on ($\J$)/DY as a function of transverse energy $E_T$.
Since there is a linear relationship between $E_T$ and $N_p$ over almost the
entire measured range, we can just multiply the measured ratios at each $E_T$
by the expected centrality dependence of the Drell-Yan process, ${E_T}^{4/3}$.
This is shown in Figure \ref{NA50jpsioverDY}. 
\begin{figure}[p]\hspace*{0.7cm}
\includegraphics[clip=,height=.44\textheight]{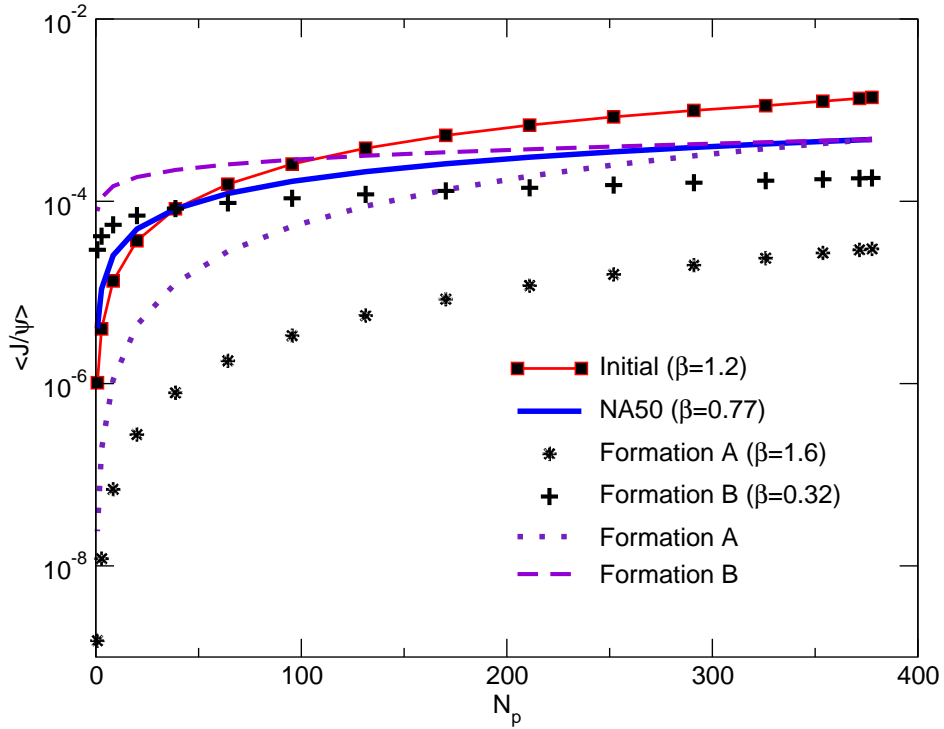}
\caption{\small Generic $\J$ production at SPS}
\label{genericspsjpsi}
\end{figure}
A power-law fit to the
resulting points yields an exponent $\beta \approx 0.77$.  This value is
significantly higher than one would predict from the generic arguments above,
since at SPS with $<\Nccbar> \, \ll 1$, $\beta \approx {4\over 3} - 
\alpha \approx 0.26$.
Figure \ref{genericspsjpsi} illustrates this feature.

As a reference, the
expected centrality dependence for initial $\J$ production is shown, which
includes the hard production process followed by normal nuclear absorption. The
parameterization of the absorption cross section is taken from \cite{Nardi}, which
leads to an effective power law exponent $\beta \approx 1.2$  The solid line
shows the centrality dependence implied by the NA50 data described above, and
is normalized such that it coincides with the initial production value for
sufficiently peripheral collisions.  The predictions of the generic quadratic formation
formula are shown by the stars (labeled A).  
Of course, the initial charm production at SPS energies is 
certainly small enough such that the linear dependence is dominant.  This
effect is shown by the plus symbols (labeled B), which include both the
quadratic and linear terms.  A single power law fit to this composite
curve yields $\beta$ = 0.32, again indicative that the linear term with its own
$\beta$ = 0.26 is dominant.
Also shown by the dotted and dashed lines
are these same curves, but normalized 
for central collisions to coincide with the ``NA50'' curve for 
ease of comparison.  It is clear that the centrality dependence implied by the
NA50 data is not reproduced by the generic expectations. 
In this respect it is fortunate that the expected magnitude
of $\J$ formation implied by the
generic arguments is likely to be insignificant
at SPS energies. 

\begin{figure}[t]
\includegraphics[height=.5\textheight]{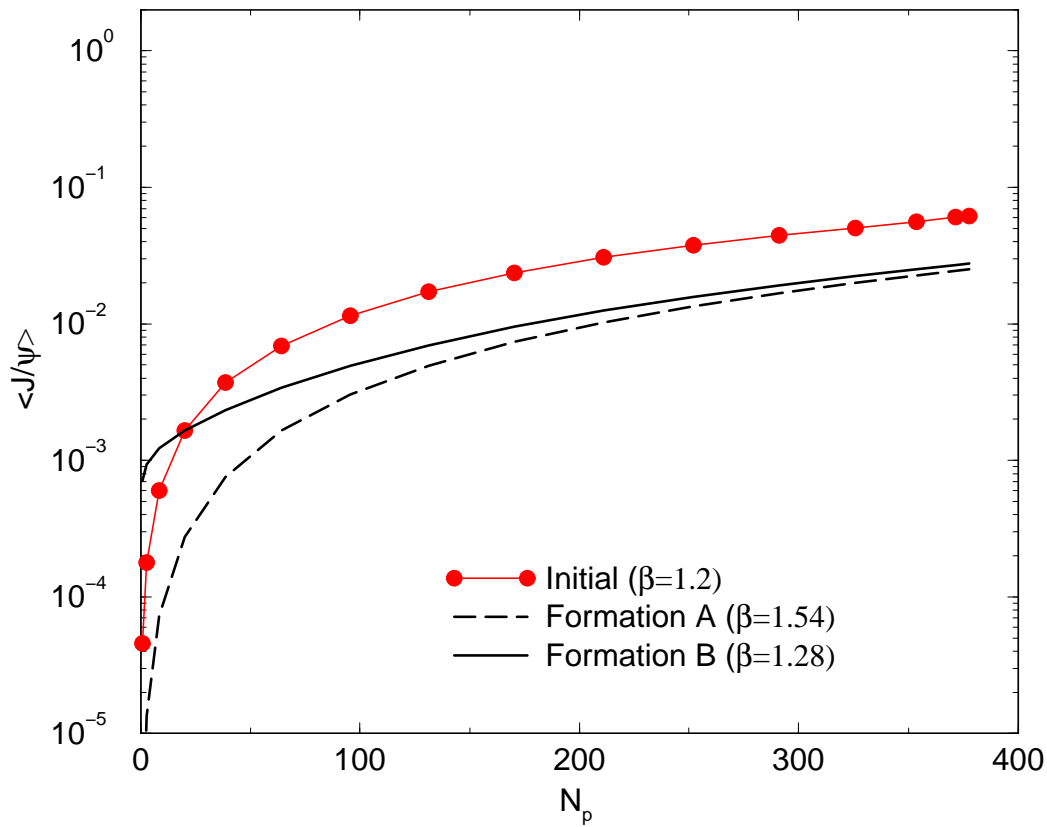}
\caption{\small Generic $\J$ production at RHIC}
\label{genericrhicjpsi}
\end{figure}
\begin{figure}[t]
\includegraphics[height=.5\textheight]{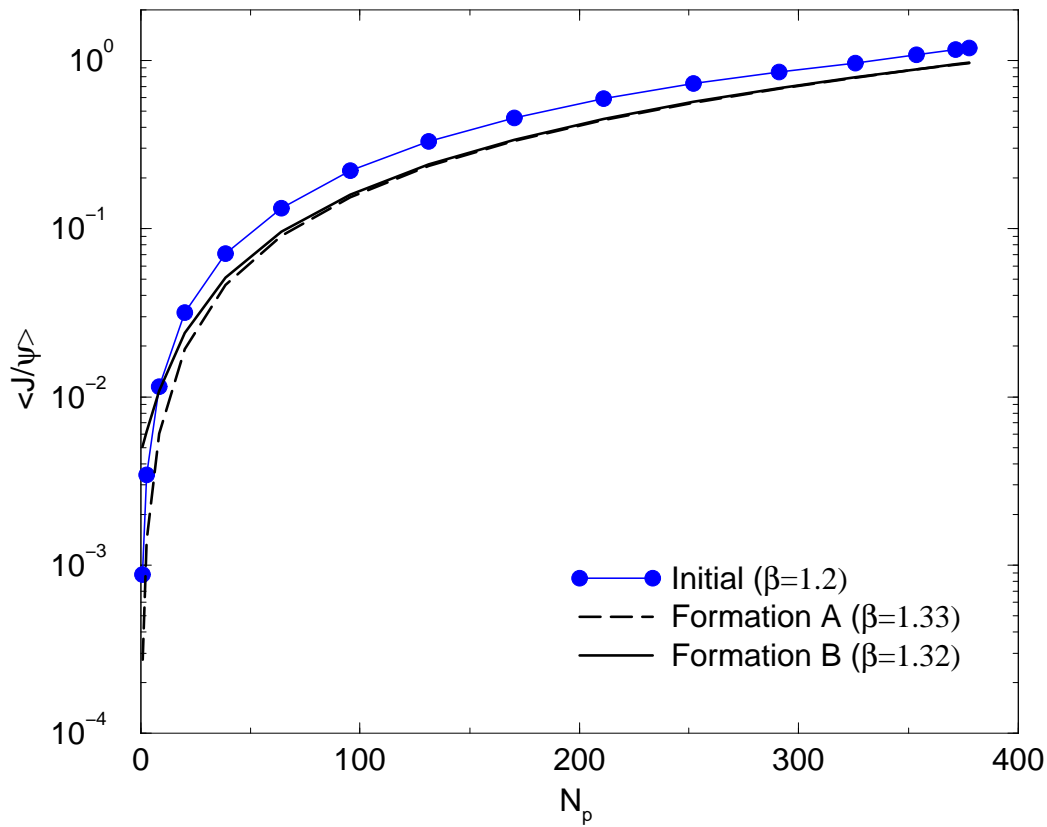}
\caption{\small Generic $\J$ production at LHC}
\label{genericlhcjpsi}
\end{figure}

The corresponding results at RHIC energy are shown in Figure \ref{genericrhicjpsi}.
Again the generic formation curves use the prefactor $\lambda$ = 1.  
Here we have used the expected charm yield of 10 pairs per central collision
at 200 GeV, but used the measured $\alpha$ value at 130 GeV.  At RHIC, one
expects to see the quadratic dependence for central collisions (Curve A) 
gradually convert to linear dependence as appropriate for the small number of
$\Nccbar$ for peripheral collisions.  Curve B shows the combination of linear
and quadratic components, which is fit by a single power $\beta$ = 1.28 (indicates
the quadratic component dominates over most of the centrality range).
Figure \ref{genericlhcjpsi} contains the same calculations at LHC energy.
Here the purely quadratic formula (B) is dominant over the entire centrality
region, since the number of charm pairs for central collisions is 
so large ($\approx 200$).  We have assumed at LHC energy that the particle
production centrality dependence has increased as appropriate for total 
domination by minijets.  If this is not correct, the $\beta$ values will be
even higher but the dominance of the quadratic term will remain.

\begin{figure}[t]
\includegraphics[height=.5\textheight]{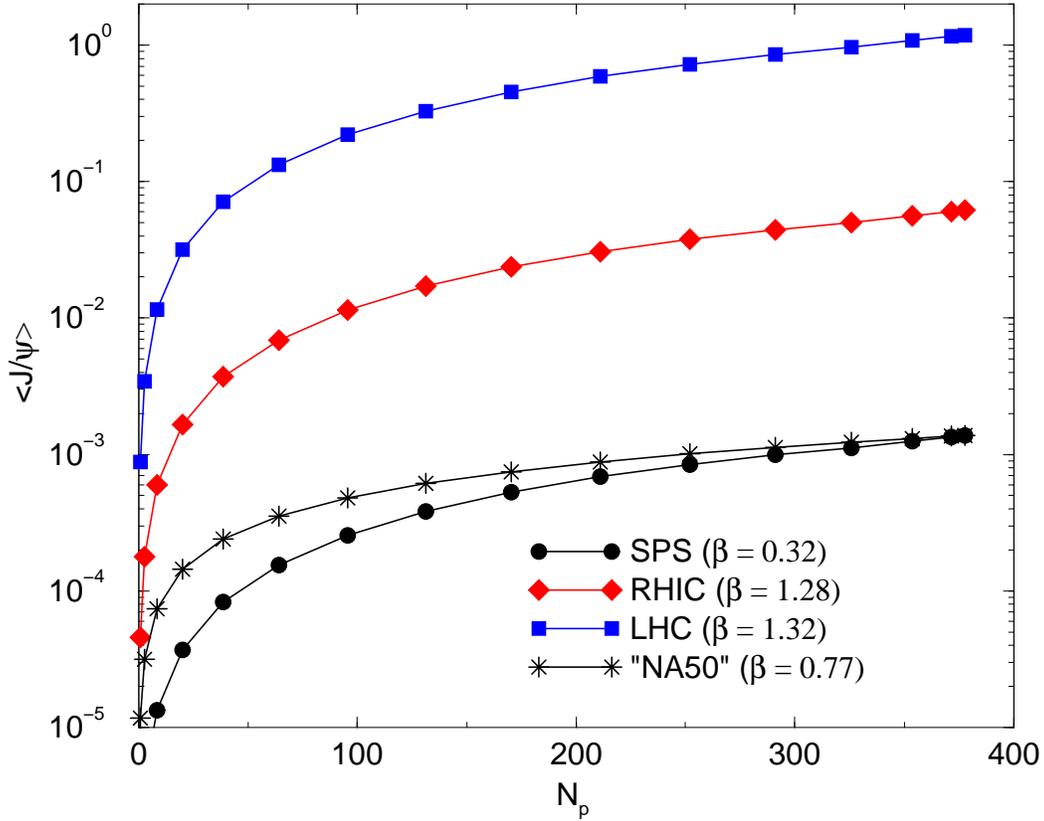}
\caption{\small Centrality dependence of generic $\J$ production with
dynamical factor $\lambda$ = 1.}
\label{genericnjpsi}
\end{figure}

We compile in Figure \ref{genericnjpsi} the expected centrality dependence
for all energies considered.  The absolute magnitudes correspond to the
prefactor $\lambda$ = 1.  Also shown is the dependence implied by the
NA50 data, which is normalized to coincide with the generic SPS curve for
the most central point.  
Certainly the centrality dependence at RHIC and LHC will
be crucial for the interpretation of any enhanced $\J$ production.


\section{STATISTICAL HADRONIZATION MODEL}

\begin{figure}[t]
\includegraphics[height=.5\textheight]{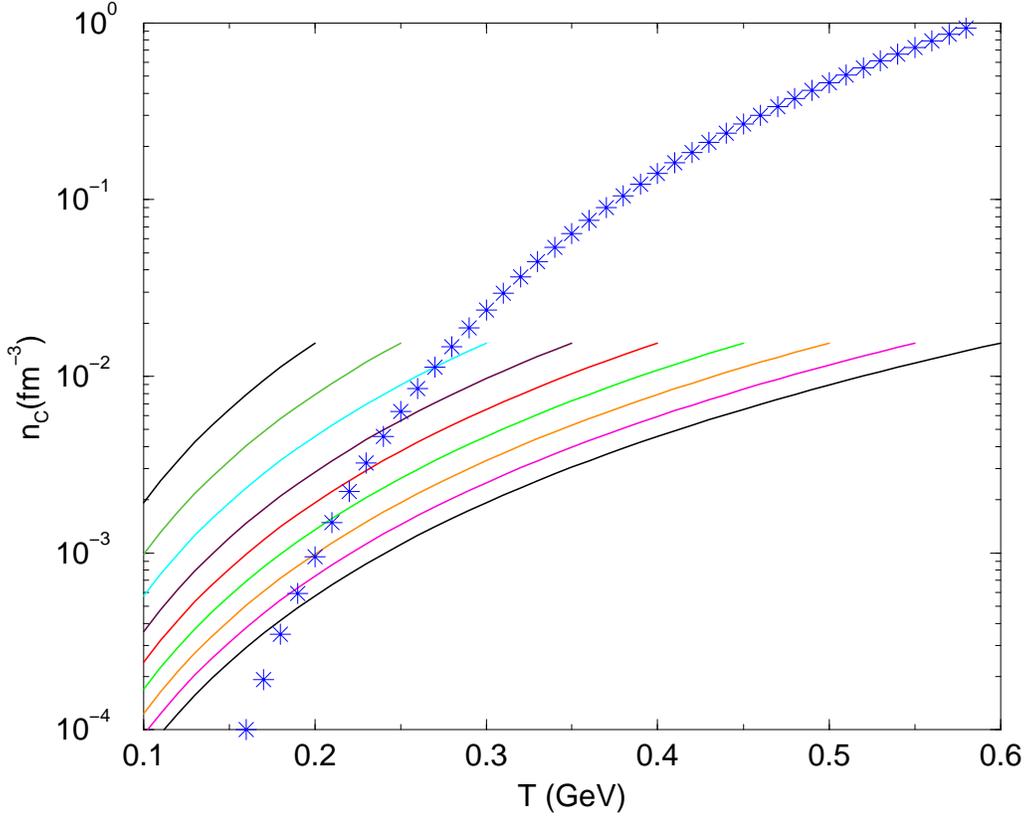}
\caption{\small Charm quark density.} 
\label{charmdensity}
\end{figure}

This model 
is motivated by the success of attempts to explain the 
relative abundances of light hadrons produced in high energy interactions in
terms of the predictions of a hadron gas in chemical and thermal equilibrium
\cite{thermalfit}.  Such fits, however, are not able to describe the
abundances 
of hadrons containing charm quarks.  This can be understood in terms of the
long time scales required to approach chemical equilibrium for heavy quarks.
However, it is expected that for high energy heavy ion collisions the
initial production of charm quark pairs exceeds the number expected at 
chemical equilibrium as determined by the light hadron abundances.  
As an illustration, we show in Figure \ref{charmdensity} the density
of charm quarks in equilibrium over a range of temperatures.  This is
compared with the charm quark density which results from distributing
the initially-produced charm quark pairs over the volume of a deconfined
region.  Each line in the figure corresponds to a different initial
temperature.  The decrease in the density as temperature decreases
is due to the expansion of the deconfined region.  (We use nominal
RHIC conditions for central collisions, $\Nccbar$ = 10, initial
volume $V_o = \pi R^2\tau_o$ with $\tau_o$ = 1 fm, and isentropic
longitudinal expansion, $VT^3$ = constant.)  One sees that at
temperatures below typical deconfinement transitions, the initial
charm densities all exceed that expected for thermal equilibrium.
(This will always happen at some finite T, since the decrease
of equilibrium density is exponential while the initial densities
only decrease due to power law volume expansion.)

The
statistical hadronization model  
assumes that at hadronization these charm quarks are distributed into
hadrons according to chemical and thermal 
equilibrium, but adjusted by a factor
$\gamma_c$ which accounts for oversaturation of charm density.  One power of
this factor multiplies 
a given thermal hadron population for each charm or anticharm
quark contained in the hadron.  Thus the relative abundance of 
$\J$ to that of D mesons, for example, will be enhanced in this model.
The enhancement factor is determined by conservation of charm, again
using the time scale argument to justify neglecting pair production
or annihilation before hadronization.

\begin{equation}
\Nccbar = {1\over 2}\gamma_c N_{open} + {\gamma_c}^2 N_{hidden},
\label{eqgcstat}
\end{equation}
where $N_{open}$ is the number of hadrons containing one charm or
anticharm quark and $N_{hidden}$ is the number of hadrons containing
a charm-anticharm pair.  (The contribution of multiply-charmed 
baryons or antibaryons are generally neglected since their large
mass leads to very small thermal densities.)  Note that the actual
particle numbers, not just the densities, are required in this
approach.  Thus the volume of the thermal system is an additional
parameter which must be included.

For most applications, $N_{hidden}$ 
in Equation \ref{eqgcstat} can be neglected compared with
$N_{open}$ due to the hadronic mass differences.  Thus the charm
enhancement factor is simply

\begin{equation}
\gamma_c = {2\Nccbar\over N_{open}}.
\end{equation}

This is easily seen to predict a quadratic dependence of the population
of hidden charm hadrons.  Using the thermal densities allows one
to calculate the prefactor $\lambda$ from the previous section.  

\begin{equation}
\NJ = {\gamma_c}^2 {\NJ}^{thermal}
\end{equation}

It is important to note at this time that $\NJ^{thermal}$ includes
the thermal population of all hidden charm states which decay into
the observed $\J$.  Since all of the individual terms are multiplied by
the same charm factor $\gamma_c^2$, the statistical hadronization
model predicts that all ratios of various hidden charm state populations
are identical to those predicted by the thermal densities alone.
It was first noted in Reference \cite{ratio} that the measured 
ratio ${\Psi^{\prime}/ \Psi}$ for heavy ion interactions at SPS 
was quite close to that expected in thermal equilibrium at 
temperature close to the deconfinement transition for sufficiently
central collisions.

We replace the one remaining factor of system volume by the 
ratio of charged particle number to density to compare with
the generic expectations in Equation \ref{eqquadratic}.

\begin{equation}
\NJ = 4 {n_{ch} n_{\J}\over {n_{open}}^2} {\Nccbar^2\over N_{ch}}
\label{eqstathad}
\end{equation}

This is the result contained in the initial formulation of the 
statistical hadronization model \cite{bms1}, where the goal was to
compare with results of the NA50 experiment on $\J$ production.  It was
soon realized \cite{bms2}, \cite{frank1} that for such an application, an
important correction must be applied to Equation \ref{eqgcstat}.  
We have tacitly assumed up to now that the thermal particle numbers
$N_{open}$ and $N_{hidden}$ have been calculated
in the grand canonical
formalism, and that they are large enough such that charm conservation 
is satisfied by the average values with fluctuations suppressed
by these large particle numbers.  However, one knows that at SPS energies
the average thermal charm numbers per collision are  much less than
unity for all collision centrality.  Even at RHIC and LHC, 
sufficiently non-central
events will always involve small particle numbers.  In these cases, 
one cannot satisfy exact charm conservation on the average, and one
must utilize the canonical formalism to calculate the thermal
particle numbers.  This is obvious in the limiting case where each
collision produces either one or zero charm quark pairs, and they can
go only into either one hidden charm hadron or one charm and one anticharm
hadron \cite{redlich1}.

There is a simple correction factor which can be applied to calculate the 
canonical particle number $N_{can}$
from the grand canonical particle number $N_{gc}$ for
a thermal system in which the conserved quantity (charm in this case)
population in baryons can be neglected (justified by large masses).
\begin{equation}
N_{can} = N_{gc} {{I_1(N_{gc})}\over {I_0
(N_{gc})}}.
\end{equation}
which just involves the modified Bessel functions $I_n$ \cite{canonical}.
One uses this expression with $N_{gc} = \gamma_c N_{open}$ to revise
Equation \ref{eqgcstat}.  Note that there is no canonical correction 
for $N_{hidden}$, which has zero total charm.

\begin{equation}
\Nccbar = {1\over 2} \gamma_c N_{open} {{I_1(\gamma_c N_{open})}\over {I_0
(\gamma_c N_{open})}} + {\gamma_c}^2 N_{hidden},
\end{equation}

In the limit of large $N_{gc}$, the ratio of Bessel functions approaches unity,
and one recovers the grand canonical result.  In the opposite limit when
$N_{gc}$ approaches zero, the ratio of Bessel functions goes to 
${1\over 2} N_{gc}$, and the solution for the charm enhancement factor is
\begin{equation}
\gamma_c \rightarrow {2\sqrt{\Nccbar}\over N_{open}}.
\end{equation}
The net effect in this limit is then just to change the dependence on
$\Nccbar$ in Equation \ref{eqstathad} from quadratic to linear.  

\begin{figure}[tb]
\includegraphics[height=.52\textheight]{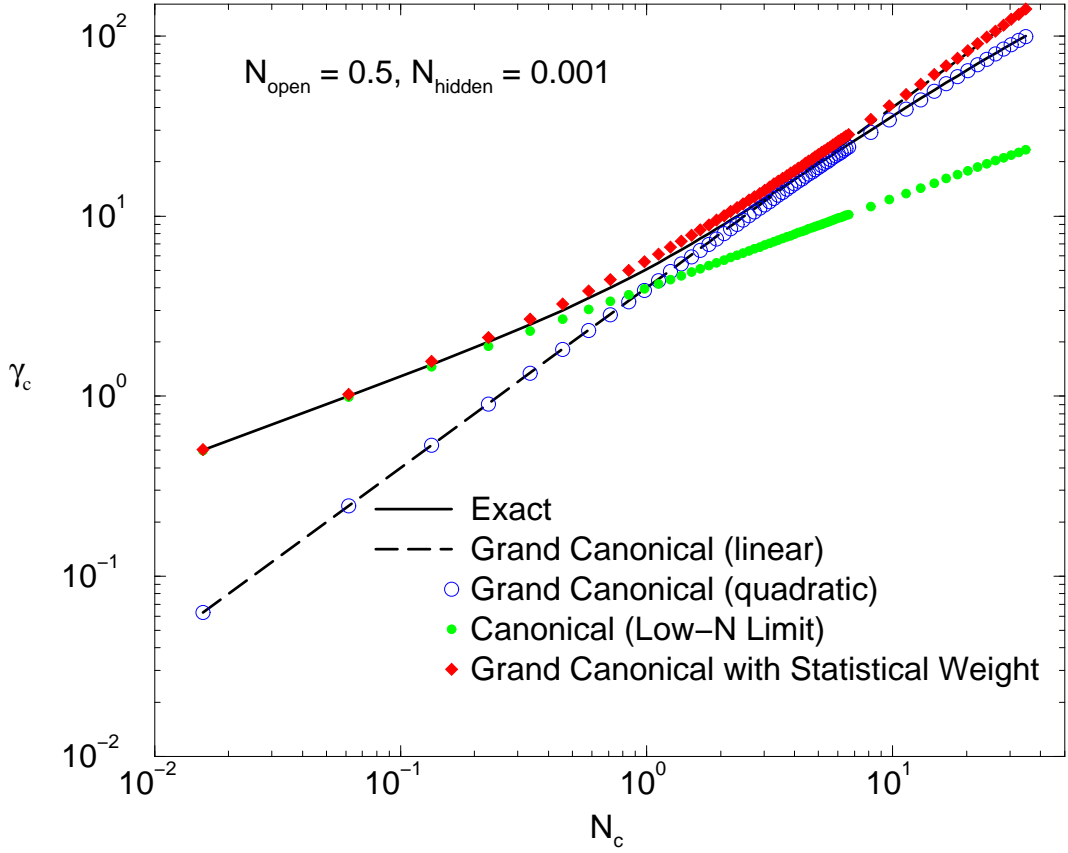}
\caption{\small Relation between charm enhancement factor and total number
of charm quarks for statistical hadronization model.}
\label{gammac}
\end{figure}

In general, a solution for $\gamma_c$ as a function of $\Nccbar$ must be
obtained numerically.  Such a solution is shown in Figure \ref{gammac}, using
some specific values of $N_{open}$ and $N_{hidden}$ for charm.  One sees
the quadratic behavior in the large $\Nccbar$ limit and also the linear
behavior in the small $\Nccbar$ limit.  
Also shown on this plot is a curve 
which follows the exact solution for Equation \ref{eqgcstat}, but with
$\Nccbar$ replaced by $(\Nccbar(\Nccbar+1))^{1\over 2}$.

It is interesting to note that this replacement allows the grand canonical 
solutions to incorporate the behavior of the canonical corrections
to an impressive degree of accuracy.  It appears that the ad hoc
substitution above could be motivated by an averaging procedure for 
$\Nccbar^2$, which has been noted previously for the two limiting
cases \cite{frank1}.  However, apparent general validity of this procedure
requires further study, and may involve the properties of the
kinetic equations which describe the approach to equilibrium.

This formalism was originally applied to the NA50 measurement of 
$\J$ production in fixed-target heavy ion collisions at the CERN SPS.
There remain uncertainties in the absolute magnitude of $\J$ yields, 
which has lead to different approaches in the literature.
One method is to assume knowledge of $\Nccbar$ from measurements in
N-N interactions scaled up to heavy ion interactions as appropriate for
a point-like process,  Then the $\J$ can be predicted, including the
centrality dependence.  The other approach takes $\Nccbar$ as a 
parameter to be fixed by the measured $\J$ yields.  In both cases 
\cite{bms2}, \cite{frank3},
a common conclusion appears to emerge.  One must require that the 
magnitude of charm production must be a factor of 3-5 greater than that 
inferred from N-N interactions, and the centrality dependence
must increase much more rapidly than the ${N_p}^{4\over 3}$ 
expected for the pQCD production process.  This conclusion may be
related to the observation of an excess of dileptons in the intermediate
mass region by NA50 \cite{bigcharm} for which one source could
be enhanced charm production.  This situation underscores the need for 
separate measurements of both total charm and charmonium production in order to
test the production mechanisms.  Experiment NA60  
is expected to provide this information at SPS.

Next we look at applications of the statistical hadronization
model at RHIC and LHC.  Since the general properties of this model 
obey the generic expectations of the previous section, we can utilize
the generic expectations for centrality dependence. The overall
magnitudes are determined by the thermal parameters (including volume), 
and are generally taken from existing thermal fits to light
hadron species.  We choose to show the ratio $<\J>$/charm, which is
less sensitive to the normalization of total charm production 
the absolute number of $\J$. 

\begin{figure}[p]
\includegraphics[clip=,height=.43\textheight]{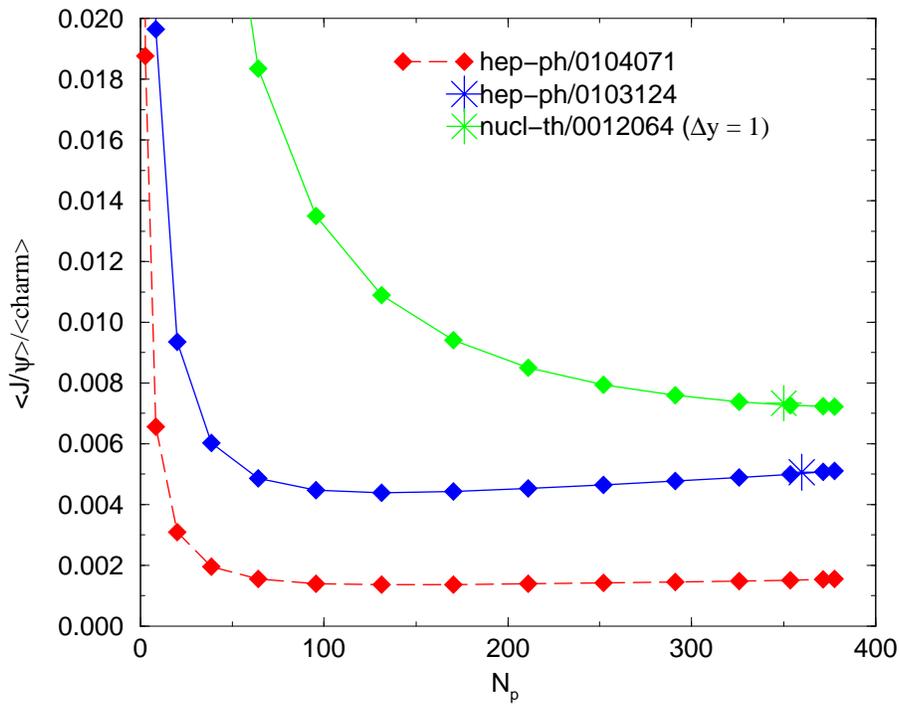}
\caption{\small Ratio $<\J>$ over initial charm at RHIC for several
applications of the statistical
hadronization model.}
\label{rhicjpsistat}
\end{figure}

Figure \ref{rhicjpsistat} shows several applications for RHIC conditions.

\begin{figure}[p]\hspace*{0.3cm}
\includegraphics[clip=,height=.43\textheight]{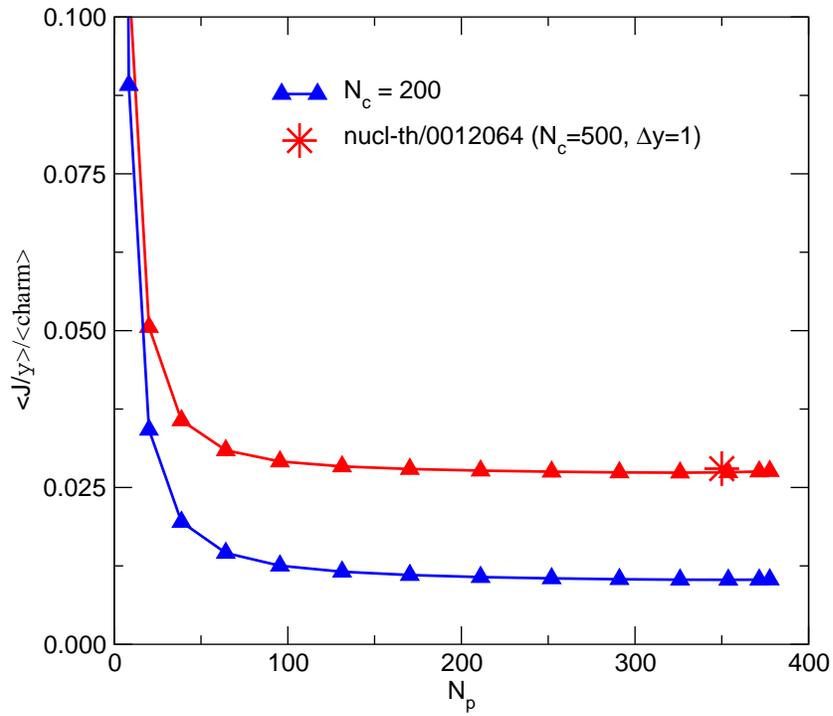}
\caption{\small Ratio $<\J>$ over initial charm at LHC for two
applications of the statistical
hadronization model.}
\label{lhcjpsistat}
\end{figure}

The centrality dependence is modeled by the number
of nucleon participants, and one sees the change in shape due to the
transition between canonical and grand canonical formalism.  The absolute
magnitudes of $<\J>$/charm are comparable with the initial production 
estimates of a fraction of a percent, indicating that this process may
overwhelm suppression for central collisions at RHIC.  The lowest 
curve is the calculation of Reference \cite{frank3}, which includes the
centrality dependence.  The next higher is from Reference \cite{rapp}
which only included the most central collision point.  The same is true
for the highest curve from Reference \cite{bms2}, which uses rapidity
intervals of width $\Delta y$ = 1 as a requirement for
the quarks to form the $\J$.  I have completed the calculation of
implied centrality dependence for these two calculations.  Note that
the highest curve exhibits a stronger rise for lower centrality events
than the others, since with total charm quark numbers limited by one unit 
of rapidity, the region which receives a substantial canonical 
ensemble correction extends further toward very central events.

Corresponding information is shown in Figure \ref{lhcjpsistat} for
LHC energy.  Here we have completed the centrality dependence
for one case considered in the literature
 \cite{bms2}, and contrast it with
the generic calculation with no constraint on rapidity interval.  
Typical magnitudes are factors of 3-5 above the corresponding
predictions for RHIC, indicating a strong enhancement of
$\J$ formation in the statistical hadronization model.


\section{KINETIC FORMATION MODEL}

In this model \cite{thews1},\cite{thews2}
, we investigate the possibility to form $\J$ directly in
a deconfined medium.  The formation will take advantage of the mobility
of initially-produced charm quarks in a spatial region of
deconfinement. Then one expects that interactions can occur between
a charm quark which was produced along with its anticharm partner
in one of the initial nucleon-nucleon collisions, and an anticharm
quark which was produced with its own charm partner in an entirely
different initial nucleon-nucleon collision.  Thus all combinations 
of a charm plus anticharm quarks in the initial $\Nccbar$ 
are allowed to participate in the formation of charmonium states.
Of course, there is an upper limit of $\Nccbar$ itself on the total
number of charmonium states which can be formed, but in practice this
limit will never be approached.  Since the rates of formation are 
quadratic in the number of {\em unbound} charm pairs, one 
anticipates that the final charmonium population will be 
approximately quadratic in the {\em initial} value $\Nccbar$.
In this respect, it then fits in with the generic expectations
previously derived based on probabilities of quark number
combinations.  However, the additional dependence of that
generic expectation on hadron production will come
about in an entirely different manner.

For the purposes of this study, we consider a physical picture
of deconfinement
in which the ``standard'' quarkonium suppression mechanism is
via collisions with free
gluons in the deconfined medium \cite{Kha}.  
The dominant formation process, 
in which a quark and an antiquark in
a relative color octet state are captured into a color singlet
bound quarkonium state and emit a color octet gluon, 
is simply the
inverse of the the breakup reaction which is responsible for 
the suppression.  It is then
an inevitable consequence of this picture of suppression that 
the corresponding formation process must also take place.

At this point one might ask about the effect of color screening
in this picture.  One view might be that the most deeply bound
quarkonium states can still exist above the deconfinement 
temperature, and that this formation mechanism (and of course
the competing dissociation mechanism) will only exist 
for temperatures above the deconfinement temperature (since we
require mobile heavy quarks)  but still below some critical
temperature $T_{screen}$ which defines the point at which the
quarkonium state can no longer exist.  In this case the new
formation mechanism will just modify the dissociation 
effectiveness, although in the case of very large numbers of
quark pairs this modification will be sufficient to actually
change the sign of the effect.  In this study we advocate an
alternate viewpoint. In this viewpoint, both the color screening
mechanism and the gluon dissociation reaction are the same
physical phenomenon, but manifest themselves in two different
limiting cases.  In the limit of very large time scales, the
screening view is appropriate, since it assumes that the 
heavy quarks are subject to a static potential which determines
the bound state spectrum.  In the limit of very small time scales,
the quarkonium states cannot be expected to follow the fluctuations
of the color fields due to the light quarks and gluons, and it
is assumed that a collisional description will be appropriate.
Work is underway \cite{sucipto} on quantifying these arguments. 
Initial results indicate that the time necessary to completely
screen away a deeply bound quarkonium state is comparable to
estimated lifetimes for a deconfined state in heavy ion collisions.

Given the above assumptions, we consider the dynamics of 
charm quark pairs and $\J$'s in a region of color deconfinement
populated by a thermal density of gluons.
The time evolution if the $\J$ population
is given by the rate equation
\begin{equation}\label{eqkin}
\frac{d\NJ}{d\tau}=
  \lambda_{\mathrm{F}} N_c\, N_{\bar c }[V(\tau)]^{-1} -
    \lambda_{\mathrm{D}} \NJ\, \rho_g\,,
\end{equation}
with $\rho_g$ the number density of gluons, $\tau$ the proper time
and $V(\tau)$ the volume of the deconfined spatial region.
The reactivities $\lambda_{F,D}$ are 
the reaction rates $\langle \sigma v_{\mathrm{rel}} \rangle$
averaged over the momentum distribution of the initial
participants, i.e. $c$ and $\bar c$ for $\lambda_F$ and
$\J$ and $g$ for $\lambda_D$.
The gluon density is determined by the equilibrium value in the
QGP at each temperature.  For simplicity, it is assumed to be 
spatially homogeneous,
as are the charm quark and $\J$ distributions.  

We enforce exact charm conservation in solving this equation, but
in practice this means that the number of charm quarks $\Nc$
and anticharm quarks $\Ncbar$ are always approximately equal
to the number of initial pairs $\Nccbar$ for the following reasons:
(a) Reactions in which charm quark-antiquark pairs
annihilate into light quarks or gluons are small since the charm
density due to initial charm is less than the thermal equilibrium
value over during most of the time; (b) Production of additional
charm quark pairs from interactions of light quarks and gluons is
negligible during the time of deconfinement \cite{letessier};
(c) Formation of other states in the charmonium spectrum is a small
fraction of initial charm (as is the fraction of $\J$ itself);
and (d) Disappearance of single charm quarks or antiquarks via
formation of open charm mesons is effectively reversed immediately
because the time scale for dissociation of these large states 
with small binding energy is typically less than a fraction of 
a fermi.

\begin{figure}[t]
\includegraphics[clip=,height=.5\textheight]{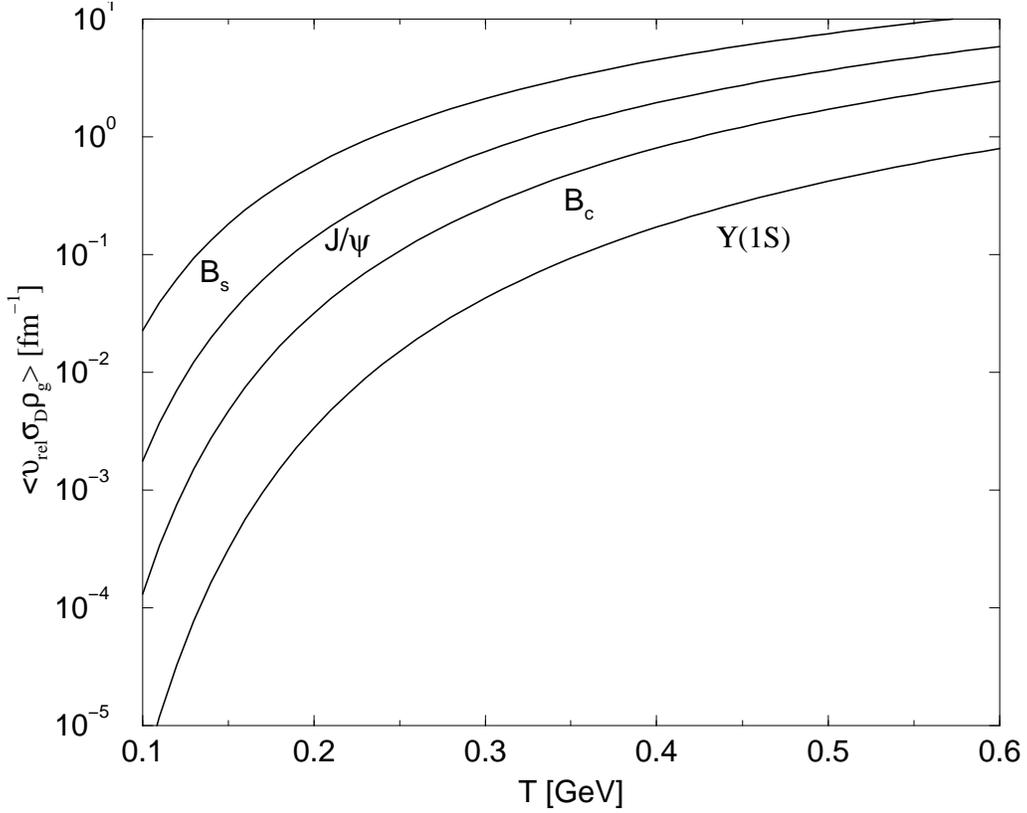}
\caption{\small Dissociation rates of various bound states of charm due to
interactions with thermal gluons.}
\label{diss}
\end{figure}

To illustrate this last point and set the relevant time scales, we
show in Figure \ref{diss} the dissociation rates for thermal gluons
on various states of quarkonium, as a function of gluon temperature.
The reaction cross section used will be discussed below.

The behavior with binding energy and spatial bound state size is 
evident.  Certainly the open charm mesons will have typical
dissociation times much less than 1 fm.  It is also clear that the
heavy quarkonium bound states have dissociation times between 1 and 
100 fm for typical QGP temperatures, making this process relevant 
during the time of deconfinement.  Note that the temperature dependence
arises from two effects.  The initial rapid increase in the
low 
temperature region starts when average gluon energies are able to
overcome the dissociation threshold, and the continued rise for
large temperature is due to the continuing increase in
gluon density.

We allow the system to undergo 
a longitudinal isentropic expansion, which fixes
the time-dependence of the volume V($\tau$) = $V_o \tau/\tau_o.$  
The expansion is taken to be isentropic, $VT^3$ = constant, which
then provides a generic temperature-time profile.

It is evident that the solution of Equation \ref{eqkin} grows quadratically
with initial charm $\Nccbar$, as long as the total $\J \ll \Nccbar$.  In
this case we can write an analytic expression

\begin{equation}
\NJ(\tau_f) = \epsilon(\tau_f) \times  [\NJ(\tau_0) +
\Nccbar^2 \int_{\tau_0}^{\tau_f}
{\lambda_{\mathrm{F}}\, [V(\tau)\, \epsilon(\tau)]^{-1}\, d\tau}],
\label{eqbeta}
\end{equation}
where $\tau_f$ is the hadronization time determined by the
initial temperature ($T_0$ is a variable parameter) and
final temperature ($T_f$ ends the deconfining phase).
The function $\epsilon(\tau_f) = 
e^{-\int_{\tau_0}^{\tau_f}{\lambda_{\mathrm{D}}\, \rho_g\,
d\tau}}$ 
would be the suppression factor in this scenario if the
formation mechanism were neglected.  During the remainder of
these calculations, we concentrate on solutions in which
the initial number of produced $\J$ is zero.  Then the additional
term due to the new formation mechanism represents the 
final number of $\J$ which result from a competition between
the formation and dissociation reactions during the lifetime of
the deconfined region.  Note that this number is always 
positive, since one cannot dissociate more bound states than
are formed.

We can then compare this additional term with what is anticipated
from the generic considerations summarized in Equation \ref{eqgeneric}.
The quadratic factor $\Nccbar^2$ is present as expected.  The
normalizing factor of $N_{ch}$ does not appear automatically.
(Remember that this factor was obtained in the statistical
hadronization model by replacing a factor of system volume
$V$ by the ratio of $N_{ch}$ over the thermal charged hadron
density.)  In the kinetic model result, the additional formation
term in Equation \ref{eqbeta} also has a volume term present.
This volume is present to account for the decreasing charm quark
density during expansion.  It also has some other essential differences.
 First, the kinetic model volume is time-dependent,
and is integrated over the duration of deconfinement.  Second, the
transverse area of the volume is determined not just by nuclear
geometry, but by the dynamics which determine over which region there will
be deconfinement.  Finally, the factor $\epsilon(\tau_f)/\epsilon(\tau)$
due to dissociation processes during deconfinement will play a role.
These differences will be seen explicitly when the centrality
dependence is considered.

For our quantitative estimates,
we utilize a cross section for the dissociation of $\J$
due to collisions with gluons
which is based on the operator product expansion
\cite{OPE},\cite{OPE2}:
\begin{equation}
\sigma_D(k) = {2\pi\over 3} \left ({32\over 3}\right )^2
\left ({2\mu\over \epsilon_o}\right )^{1/2}
{1\over 4\mu^2} {(k/\epsilon_o - 1)^{3/2}\over (k/\epsilon_o)^5},
\label{eqsigma}
\end{equation}
where $k$ is the gluon momentum, $\epsilon_o$ the binding energy,
and $\mu$ the reduced mass of the quarkonium system.  This form
assumes the quarkonium system has a spatial size small compared
with the inverse of $\Lambda_{QCD}$, and its bound state
spectrum is close to that in a nonrelativistic Coulomb potential.
These assumptions are somewhat marginal for the charmonium 
spectrum, but should be better satisfied for the upsilon states.

The magnitude of the cross section is controlled just by the
geometric factor ${4\mu^2}$, and its rate of increase
in the region just above threshold is due to phase space and
the p-wave color dipole interaction.
This same cross section is utilized with detailed balance factors
to calculate the primary formation rate for the capture of
a charm and anticharm quark into the $\J$.

We use parameter values for thermalization
time $\tau_0$ = 0.5 fm, initial volume $V_0 = \pi R^2\tau_0$ with
R = 6 fm, deconfinement temperature $T_f$ = 150 MeV, 
and a wide range of initial temperatures 200 MeV $< T_0 <$ 600 MeV.
We begin by showing some results which assume also thermal charm quark
momentum distributions, but this will be generalized later.

\begin{figure}[p]
\includegraphics[height=.43\textheight]{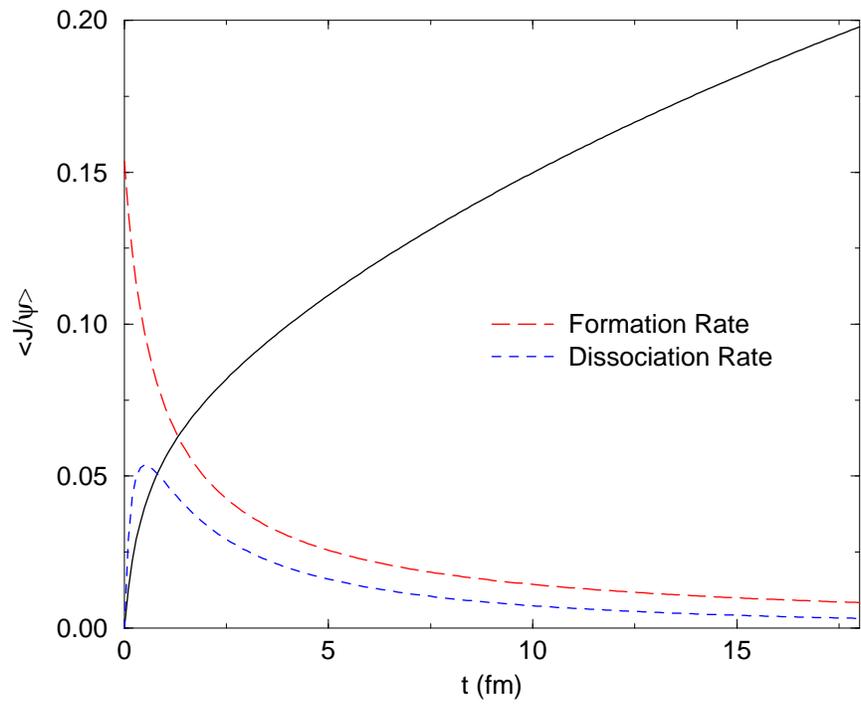}
\caption{\small Time dependence of formation and dissociation rates
and total $<\NJ>$ at RHIC energy.}
\label{timeplot}
\end{figure}

Shown in Figure \ref{timeplot} is one typical time evolution of
$\J$ taken from our numerical solutions.  Also shown are 
the time dependence of the formation and dissociation reactions.
\begin{figure}[p]\hspace*{0.3cm}
\includegraphics[clip=,height=.44\textheight]{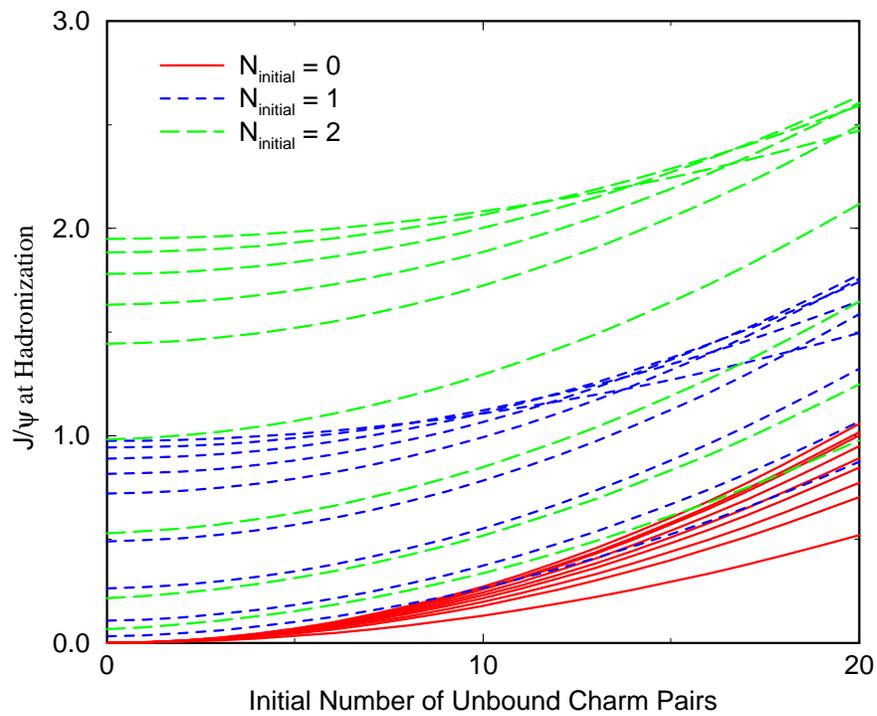}
\caption{\small Quadratic dependence of $<\J>$ as a function of total initial
charm.}
\label{quadratic}
\end{figure}
falls due to the decrease in charm quark density with the
expanding volume.  The dissociation rate starts out at zero
since there are no initial $\J$'s, but then jumps up to follow
the $\J$ population.  This rate also eventually decreases due 
to the drop in average gluon energy related to the falling
temperature.  The net number of $\J$ formed continues to increase
with time as the formation and dissociation rates both slowly
decrease.   One might wonder at what point one would reach
equilibrium, where the two rates would cancel exactly.  The answer
is that in this case we have controlled the temperature 
by external means, and equilibrium is never attained during the
lifetime of the deconfinement.  If we had assumed a constant
temperature and volume, there would of course be an equilibrium
point beyond which the $\J$ population would become constant.

Figure \ref{quadratic} shows the final $\J$ population as 
a function of $\Nccbar$ for a range of values which include
that expected for central collisions at RHIC.

For these calculations, we also allowed the initial $\J$ number to
be nonzero.  The solid, short dashed, and dashed lines correspond
to $\NJ(\tau_o)$ = 0, 1, and 2, respectively.  Variation within
these line types is from variation of the initial temperature
parameter $T_o$, which also controls the volume expansion and
lifetime.  One sees the expected quadratic dependence on $\Nccbar$ for
all parameter choices.  The solid curves are the quantity of main
interest, where the initial $\J$ is zero.  At the expected $\Nccbar$ = 10,
one sees final $\J$ in the range of 0.1 to 0.3 per central collision. This
is at or above the number one would expect from 10 initial ccbar pairs
in N-N interactions, and suggests that the formation process will
be competitive at RHIC energy.  The two sets of dashed curves show
how the dissociation of the initial population adds to the final
number, and the expected temperature dependence of this
dissociation.

\begin{figure}[t]
\includegraphics[clip=,height=.5\textheight,angle=-90]{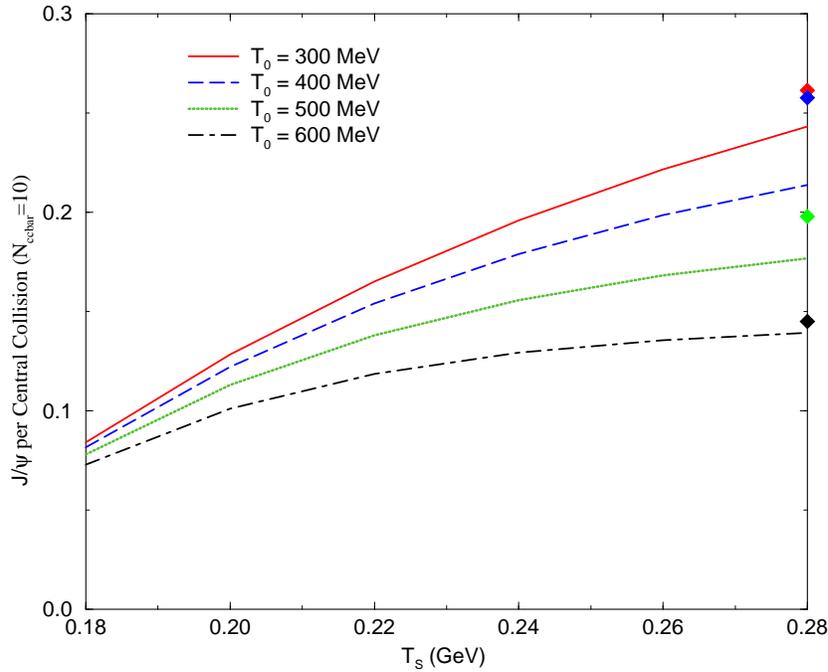}
\caption{\small Predictions of kinetic model for
$<\J>$ with maximum temperature bounded by screening.}
\label{jpsiscreen}
\end{figure}

\begin{figure}[p]
\includegraphics[clip=,height=.445\textheight]{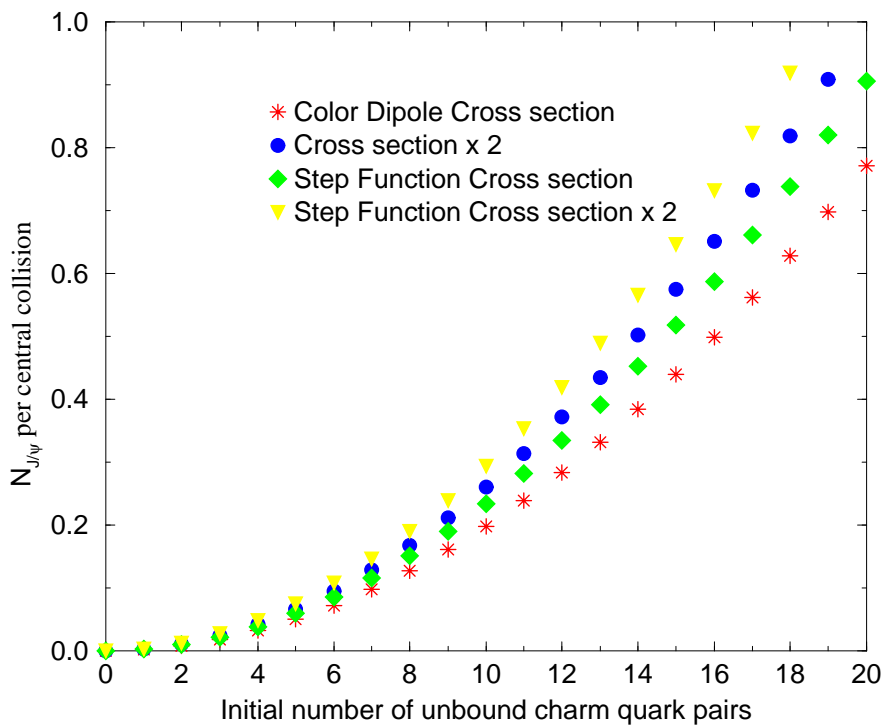}
\caption{\small Variation of formation results with input cross section.}
\label{jpsicross}
\end{figure}

\begin{figure}[p]
\includegraphics[clip=,height=.445\textheight]{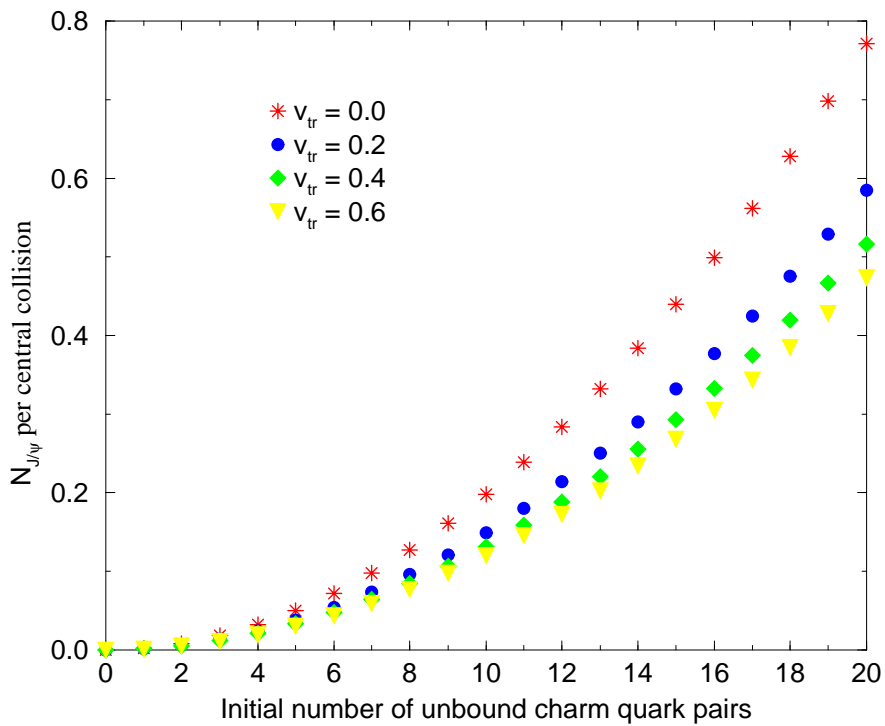}
\caption{\small Variation of formation results with nonzero transverse
expansion.}
\label{transverse}
\end{figure}

At this point we digress somewhat to examine the effects of 
varying some of our parameters or assumptions.
\begin{itemize}
\begin{figure}[t]
\includegraphics[clip=,height=.5\textheight,angle=-90]{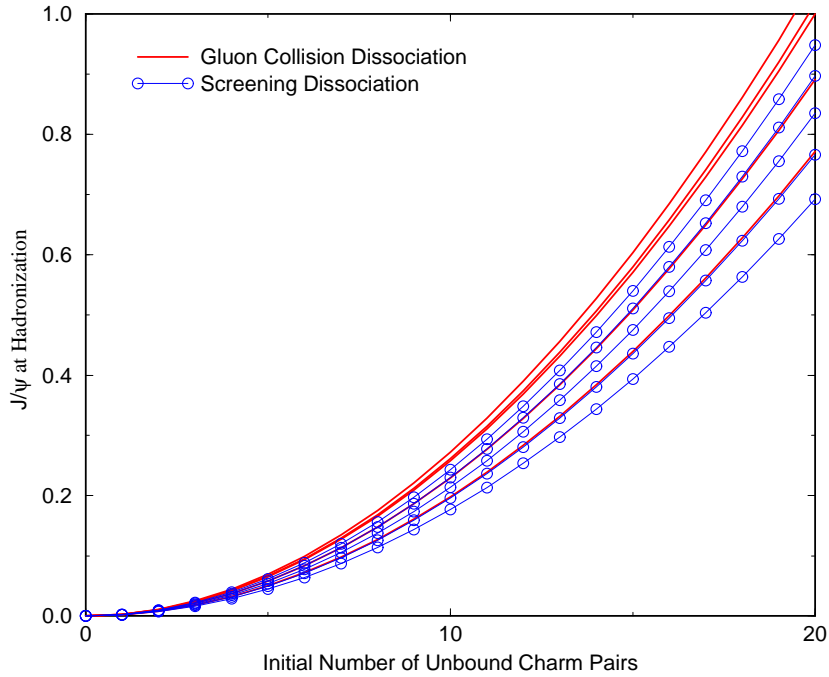}
\caption{\small Comparison of screening and collisional dissociation
scenarios.}
\label{jpsiscreen280}
\end{figure}
\item
 In our model of a deconfined region, we have used the vacuum
values for masses and binding energy of $\J$, and assumed that the
effects of deconfinement are completely included by the dissociation
via gluon collisions.  For a complementary viewpoint, we have also
employed a deconfinement model in which the $\J$ is completely
dissociated when temperatures exceed some critical screening
value $T_s$.  Below that temperature, the new formation mechanism
will still be able to operate, and we use the same cross sections
and kinematics.  

We find that for $T_s$ = 280 MeV, the final $\J$ population
is approximately unchanged.  This behavior is shown in 
Figure \ref{jpsiscreen280}.  The solid lines are our previous 
yield curves using gluon collision dissociation for various
$T_o$, and the circles show the corresponding results with a
screening cutoff temperature $T_s$ = 280 MeV.  One finds that 
for lower values of $T_s$ the formation mechanism produces 
fewer $\J$, since the most effective formation period is 
during the initial times when the total volume is small and the
corresponding charm quark densities are large.  Shown in Figure 
\ref{jpsiscreen} is the variation of the final yields with 
$T_s$, for various temperature-time profiles.  One sees that one
could get reductions in formation up to factors of 2 or 3 in this
scenario.

\item
   The validity of the cross section used assumes strictly
nonrelativistic bound states, which is somewhat marginal for the
$\J$.  All of the existing alternative models predict larger
values for this cross section.
If we arbitrarily
increase the cross section
by a factor of two, or alternatively set the cross section to its
maximum value (1.5 mb) at all energies,
we find an increase in the final
$\J$ population of about 15\%.  This occurs because the kinetics
always favors formation over dissociation, and a larger cross section
just allows the reactions to approach completion more easily within
the lifetime of the QGP.  This behavior is shown in Figure \ref{jpsicross}.

\item
     A nonzero transverse expansion will be expected at some level,
which will reduce the lifetime of the QGP and reduce the
efficiency of the new formation mechanism.  We have calculated
results for central collisions with variable transverse velocity, and
find a decrease in the equivalent $\lambda$ factor of about 15\% for each
increase of 0.2 in the transverse velocity.  This behavior is shown
in Figure\ref{transverse}.

\end{itemize}


The new formation mechanism exhibits a significant sensitivity to
the charm quark momentum distribution, as might be anticipated.  In 
the calculation of the formation reactivity $\lambda_F$, momenta of
any given pair will determine the reaction energy which in turn 
determines the effective value of the cross section.  In addition, the
charm quark energies enter into the formation rate through the usual
formulas for non-collinear reactions.  Thus we consider a large range of
possibilities for these distributions.  At one extreme, we model the
distribution to simulate the initial production distribution from the
pQCD calculations \cite{hardprobes1}.  The transverse momentum $p_T$
distribution is taken to be Gaussian with a width of 1 $GeV^2$.
The rapidity dependence is taken as flat, with the width of the
plateau $\Delta$y = 4.  We then allow for thermalization and energy
loss processes in the deconfinement region by reducing the $\Delta$y,
terminating at one unit.  This corresponds approximately to the
charm quark momentum distribution if complete thermalization
were attained for all charm quarks in the central rapidity unit.

\begin{figure}[t]
\includegraphics[clip=,height=.55\textheight]{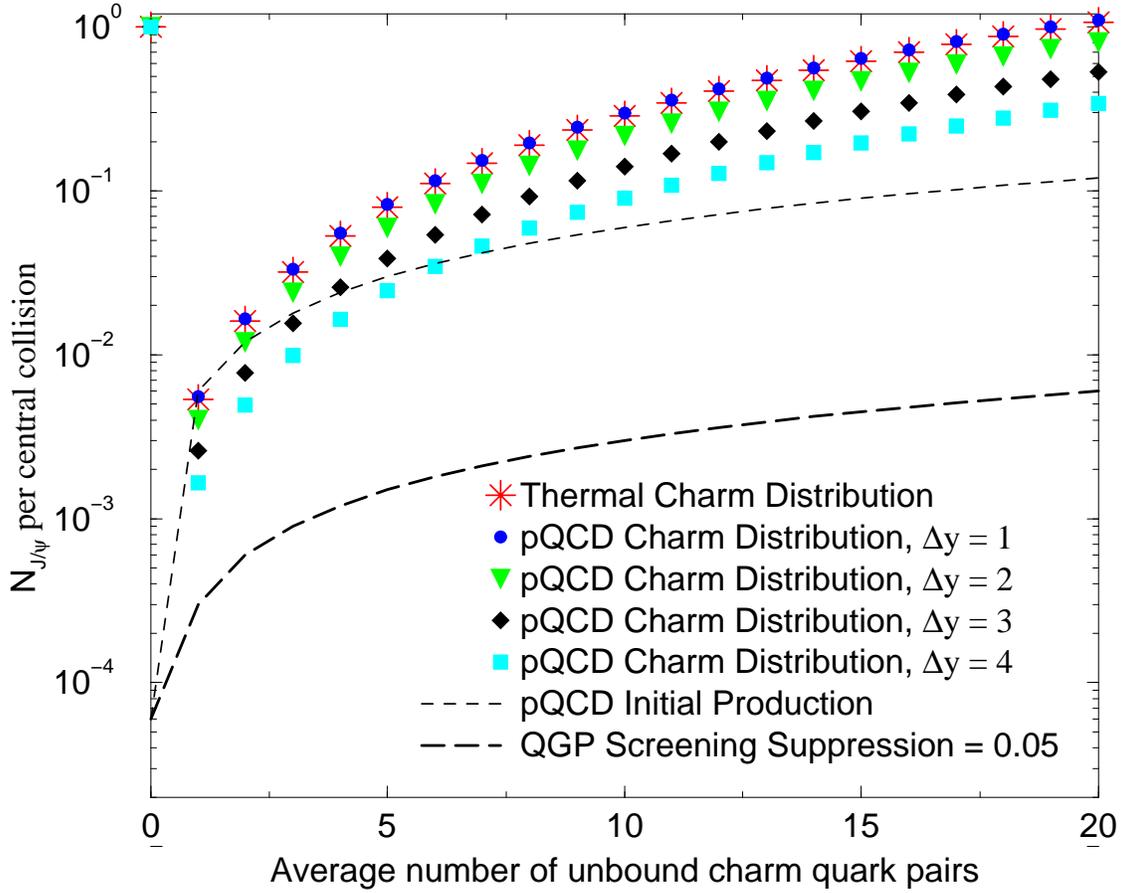}
\caption{\small Predictions of kinetic model  variation of
$<\J>$ due to charm momentum distributions.}
\label{charmdist}
\end{figure}

The results of the calculations are shown in Figure \ref{charmdist}.
Shown are curves for final $\NJ$ for central RHIC collisions as a function
of $\Nccbar$.  All of the curves for various charm quark momentum
distributions use zero initial $\J$ and an initial temperature 
$T_o$ = 300 MeV to specify the gluon density and the lifetime of
the deconfined region.  Also shown for reference is the number of 
{\em initial} $\J$ which would be produced without any 
nuclear or final state effects, which we approximate by 
$0.01 \Nccbar$ \cite{hardprobes2}.
The lowest curve just scales the initial production curve down by a
factor 0.05, which is a typical suppression factor for RHIC conditions
estimated from applications of suppression-only deconfinement models 
\cite{Vogt}.  It is seen that over even this wide range of
charm quark momentum distributions, the formation predictions are above
the initial production estimate, thus indicating an {\em enhancement}
in $\J$.  (This enhancement is even more extreme if the base is taken
to be the suppressed initial number without any formation mechanism.)

\begin{figure}[t]
\includegraphics[height=.55\textheight,angle=-90]{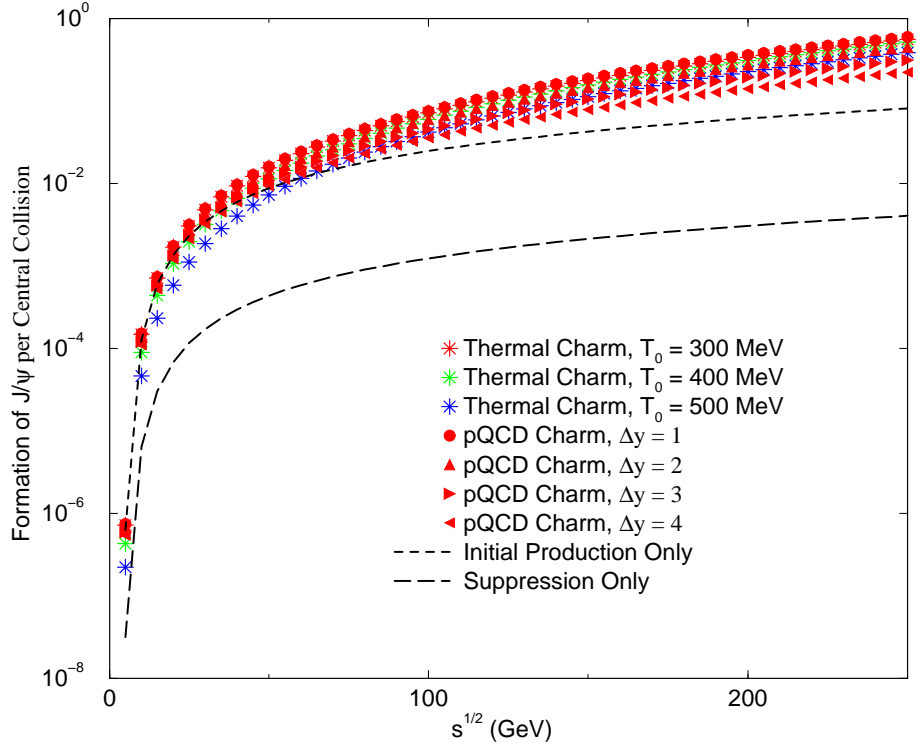}
\caption{\small Predictions of kinetic model  variation of
$<\J>$ as a function of $(\sqrt s)_{NN}$ for RHIC conditions.}
\label{energydist}
\end{figure}

We also show Figure \ref{energydist} the equivalent energy  dependence of these results, 
using RHIC conditions for the deconfinement properties but replacing
$\Nccbar$ by $\sqrt{s}$ from the pQCD expectations.

The centrality dependence will provide another prediction of
this model.  The most significant effect of varying centrality is
to sweep through a range of $\Nccbar$ from the centrality
dependence of initial production as modeled by $T_{AA}(b)$, which we
will change to the participant number dependence $T_{AA}(N_p)$.
However, a number of other components of the kinetic model
will change with centrality, through the effect of nuclear
geometry on the initial conditions and spatial properties of
the deconfined region.

The initial temperature $T_o$ is expected to decrease with increasing
 impact parameter $b$, due to a decrease in the local energy density.
We model the energy density in terms of the local density of 
participant nucleons in the transverse plane $n_w(b, s=0)$, which
is shown in Figure \ref{impactparameter}.  The dependence is then

\begin{equation}
T_o(b) = T_o(0) [n_w(b, s=0)/n_w(0, s=0)]^{1\over 4}
\end{equation}

This dependence however is not very significant until the 
very peripheral region is reached.

One also needs the initial transverse size of the deconfined region.
We model this area as the ratio of the participant number to the local
density of participants, effectively using the fall-off of density to
define the effective area within which the total number of participants
would result if their density remained at its maximum value.  Again 
this is not an absolute statement, since we normalize all of the 
centrality-dependent quantities to their values at b = 0.

\begin{equation}
A_T(b) = A_T(b = 0) [N_w(b) n_w(0, s=0)/N_w(0) n_w(b, s=0)]
\end{equation}

The results are most conveniently displayed in terms of the ratio
$\NJ/\Nccbar$, which eliminates the trivial dependence on one power of
$\Nccbar$ expected in any physical mechanism for production of $\J$.
The centrality
dependence of this ratio should be proportional
to $N_p^{4/3-\alpha}$, where now $\alpha$ contains the net effect of
the kinetic model centrality dependence.

\begin{figure}[p]
\includegraphics[clip=,height=.445\textheight]{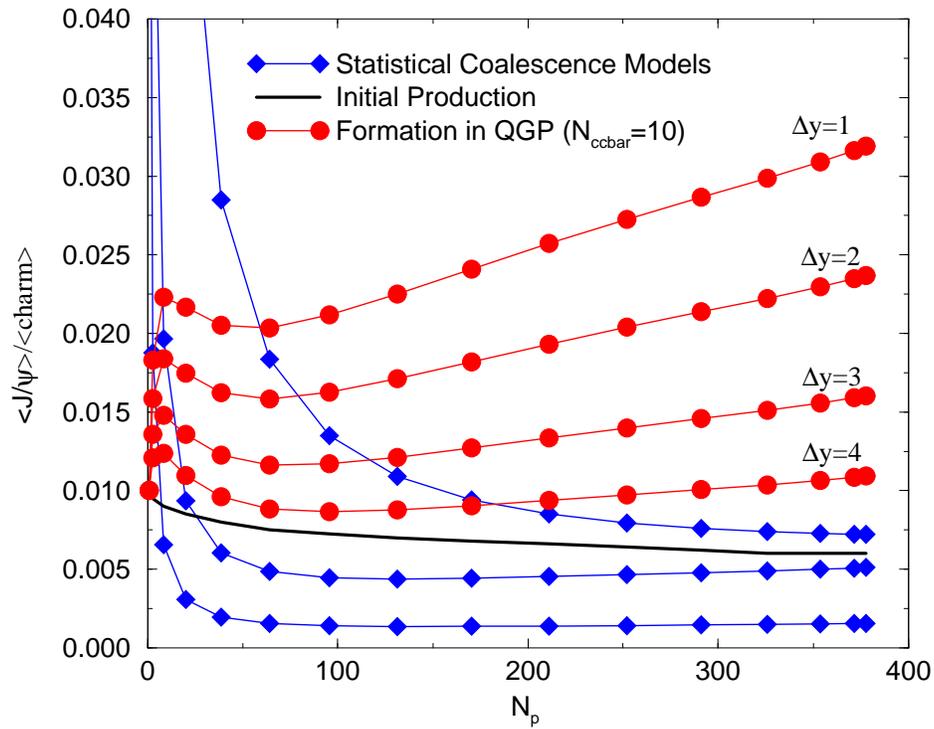}
\caption{\small Predictions of kinetic model
for $<\J>$ over initial charm at RHIC energy.}
\label{rhicjpsiall}
\end{figure}

The results for RHIC conditions are shown in Figure \ref{rhicjpsiall}.
The canonical $\Nccbar$ = 10 value for central collisions is
assumed, and required to vary with centrality as previously
discussed. 
One sees the increase with large centrality due to the quadratic behavior of 
formation, which is a characteristic signature of this type of
mechanism.  It contrasts nicely with the initial production curve, 
which has the opposite behavior due to nuclear absorption of the
initially-produced $\J$.  (The corresponding suppression-model curves
would decrease even more rapidly with increasing centrality).
The behavior for very peripheral collisions is probably an artifact
of the procedure to determine the volume of the deconfinement region, 
and should not be taken at face value.
One also sees the range of absolute values which result from 
variation of the charm quark momentum distribution.  For sufficiently
central collisions, all of these distributions predict final $\J$
which exceed the initial production, i.e. enhancement.

Also shown for comparison are the statistical hadronization
model calculations previously presented in Figure \ref{rhicjpsistat}. 
The magnitudes are somewhat lower for the cases considered, but
probably can be made compatible with some variation in parameters.
The primary difference would appear to be the sharp increase for
peripheral collisions due to the canonical corrections for
small particle numbers.

\begin{figure}[p]
\includegraphics[clip=,height=.445\textheight]{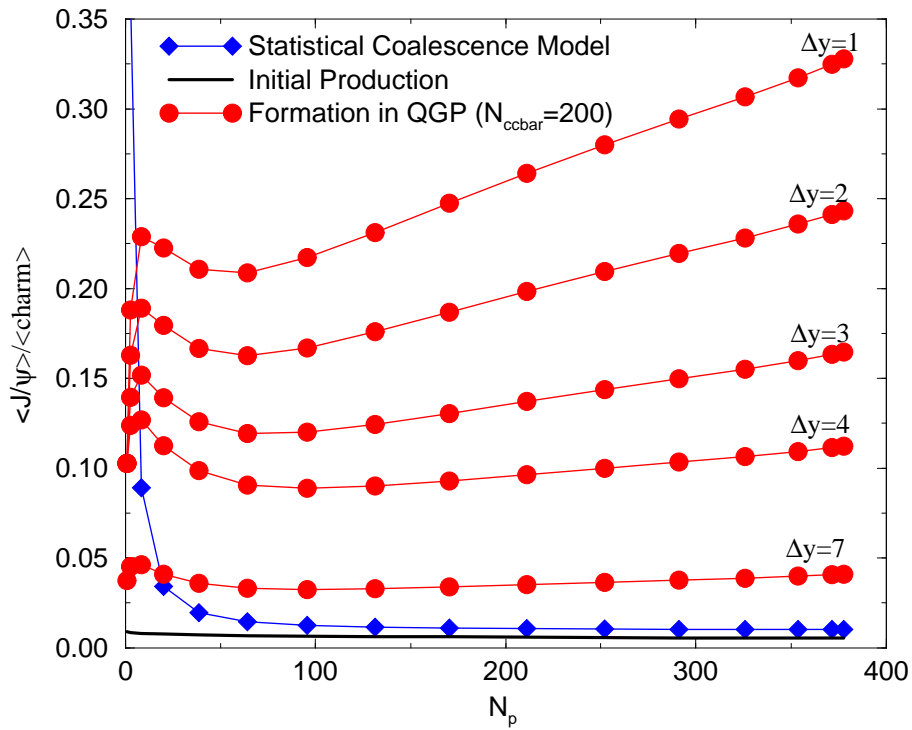}
\caption{\small Predictions of kinetic model for
$<\J>$ over initial charm at LHC energy.}
\label{lhcjpsiall}
\end{figure}

Figure \ref{lhcjpsiall} shows the corresponding predictions for
LHC.  Here we have used $\Nccbar$ = 200 for a central Pb-Pb collision
at LHC.  The range of charm quark momentum distributions has been
increased to include up to $\Delta$y = 7 to account for the
increased energy.  It is clear that the absolute magnitudes of 
$\NJ/\Nccbar$ are larger than at RHIC. This is due to the increase in
$\Nccbar$ as contained in the quadratic dependence of the 
new formation mechanism.



\section{SUMMARY}
Expectations based on general grounds for enhanced formation of heavy 
quarkonium in relativistic heavy ion collisions have been verified in
two different models.  In particular, one expects at RHIC and LHC to see
an enhancement in the heavy quarkonium formation rate, even when compared
to unsuppressed production via elementary nucleon-nucleon collisions in
vacuum.  The magnitude of this effect is expected to grow with 
the centrality of the heavy ion collision, just opposite to the predictions
of various suppression scenarios.
The physics bases for these models, however, are
quite distinct.  Their differences should manifest themselves in details of the
magnitudes and centrality dependence.  In this regard, it is essential
to have a simultaneous measurement of open flavor production to serve
as an unambiguous baseline.


\begin{theacknowledgments}
This work has been supported by U.S. DOE grant FG03-95SC306700 and
by NSF grant INT-0003184.
\end{theacknowledgments}


\bibliographystyle{aipproc}   

\bibliography{thewsproc}

\IfFileExists{\jobname.bbl}{}
 {\typeout{}
  \typeout{******************************************}
  \typeout{** Please run "bibtex \jobname" to obtain}
  \typeout{** the bibliography and then re-run LaTeX}
  \typeout{** twice to fix the references!}
  \typeout{******************************************}
  \typeout{}
 }

\end{document}